\documentclass[reprint,floatfix,notitlepage]{revtex4-1}
\usepackage[T1]{fontenc}
\usepackage[a4paper]{geometry}
\usepackage{hyperref}
\usepackage{graphicx}
\usepackage{color}
\usepackage{multirow}
\usepackage[caption=false]{subfig}
\makeatletter
\@ifundefined{date}{}{\date{}}
\makeatother

\usepackage{amsmath}
\usepackage{mathtools}
\usepackage{physics}
\usepackage{bbm}
\usepackage{pifont} 

\usepackage{soul} 

\newcommand{\ncmd}{\newcommand}
\ncmd{\nn}{\nonumber}
\ncmd{\half}{\frac{1}{2}}
\ncmd{\mbf}[1]{\bs{#1}}
\ncmd{\gam}{\gamma}
\ncmd{\sig}{\sigma}
\ncmd{\pha}{\alpha}
\ncmd{\lam}{\lambda}
\ncmd{\dl}{\delta}
\ncmd{\kap}{\kappa}
\ncmd{\eps}{\epsilon}
\ncmd{\Lam}{\Lambda}
\ncmd{\Gam}{\Gamma}
\ncmd{\Dl}{\Delta}
\ncmd{\Ups}{\Upsilon}
\ncmd{\Om}{\Omega}
\ncmd{\om}{\omega}
\ncmd{\veps}{\varepsilon}
\ncmd{\vphi}{\varphi}
\ncmd{\vtheta}{\vartheta}
\ncmd{\tw}{\text{w}}
\ncmd{\pll}{\parallel}
\ncmd{\mc}{\mathcal}
\ncmd{\mf}{\mathfrak}
\ncmd{\bs}{\boldsymbol} 
\ncmd{\trans}[1]{{#1}^\intercal}
\ncmd{\note}[1]{{\color{red}{\ding{168} #1}}}
\ncmd{\eq}[1]{Eq. \eqref{#1}}
\ncmd{\fig}[1]{Fig. \ref{#1}}
\ncmd{\suppl}{\note{`Supplementary Information'}}
\ncmd{\bc}{\text{BC}}
\ncmd{\pd}[1]{\partial_{#1}}

\definecolor{blue2}{rgb}{0.2, 0.2, 0.6}
\definecolor{blue3}{rgb}{0.16, 0.32, 0.75}

\ncmd{\para}[1]{\paragraph*{{\color{black}{\bf #1:}}} }

\ncmd{\edit}[1]{{\color{blue}{ #1}}}
\ncmd{\qs}[1]{{\color{red}{ #1}}}
\ncmd{\qsnote}[1]{{\color{magenta}{ #1}}}
\ncmd{\cs}[1]{{\color{cyan}{ #1}}}

\ncmd{\yw}[1]{{\color{green}{#1}}}

\begin{document}

\title{Fully nonequilibrium Hall response from Berry curvature
}
\author{Shouvik Sur$^{1}$}
\author{Lei Chen$^{1,2}$}
\author{Yiming Wang$^{1}$}
\author{Chandan Setty$^{1,3,4}$}
\author{Silke Paschen$^{5}$}
\author{Qimiao Si$^{1}$}
\affiliation{$^{1}$ Department of Physics and Astronomy, Extreme Quantum Materials Alliance, Smalley-Curl Institute, 
Rice University, Houston, Texas 77005, USA}
\affiliation{$^{2}$ Department of Physics and Astronomy, Stony Brook University, Stony Brook, NY 11794, USA}
\affiliation{$^{3}$ Department of Physics and Astronomy, Iowa State University, Ames, Iowa 50011, USA}
\affiliation{$^{4}$ Ames National Laboratory, U.S. Department of Energy, Ames, Iowa 50011, USA}
\affiliation{$^{5}$ Institute of Solid State Physics, Vienna University of Technology, Wiedner Hauptstr. 8-10, 1040
Vienna, Austria}

\date{\today}
\begin{abstract}
In topological materials, 
 Berry curvature leads to intrinsic Hall responses.
Focusing on time-reversal symmetric systems with 
broken inversion symmetry, 
a spontaneoous (zero magnetic field) Hall effect is expected to develop under an applied electric field.
Motivated by recent developments in Weyl-Kondo semimetals,
here we advance a fully nonequilibrium (FNE) Hall response 
due to the Berry curvature.
In particular, we show that, while the 
spontaneous 
Hall current 
is quadratic in the 
previously described regime of weak electric field,
due to the contribution from the dipole moment of the Berry curvature, the FNE Hall response for non-perturbative electric fields is not controlled by the Berry curvature dipole.
Remarkably, the 
FNE Hall response 
resembles what happens in systems that break the microscopic time-reversal symmetry.
We  illustrate the universality of these results 
by comparing them with their counterparts in systems with any higher-multipole of the Berry curvature.
The implications of our results for the understanding of strongly correlated topological semimetals are discussed.
\end{abstract}

\maketitle

\twocolumngrid

\para{Introduction} 
The interplay between topology and strong correlations is expected to produce novel quantum phases, and this is especially so in metallic settings~\cite{Pas21.1,Wit14.1}.
In recent years, Weyl-Kondo semimetals (WKSMs) have provided theoretical~\cite{lai2018weyl,chen2022topological} and experimental~\cite{dzsaber2017kondo,dzsaber2021giant} 
realizations of strongly correlated topological semimetals.
They arise in Kondo systems that preserve the time-reversal symmetry (TRS) while breaking the inversion symmetry.
An important signature of the WKSMs comes from a 
spontaneous (i.e., zero magnetic field) Hall response~\cite{dzsaber2021giant,grefe2020weyl}.

The spontaneous Hall response is distinct from the anomalous Hall 
effect, though both originate from the  Berry curvature supported by the Bloch wavefunctions.
The latter, representing a cornerstone for topological response~\cite{jungwirth2002anomalous,  fang2003anomalous, yao2004first, thonhauser2005orbital,xiao2006berry,xiao2010,nagaosa2010anomalous},
arises in topological materials
that lack TRS. By contrast, the spontaneous Hall effect develops in
systems that preserve TRS but break inversion symmetry,  
with the Hall current being 
a non-linear function of 
the electric field ($E$)~\cite{deyo2009, sodemann2015,low2015, zhang2023}.
The non-linear dependencies arise  from  higher-order  moments of the Berry curvature~\cite{sodemann2015}  {due to an equal number of current carrying states supporting opposite Berry curvature densities}, and a correspondence between the order of the moment and the leading scaling of the Hall current with $E$ in the weak-field limit has been established~\cite{zhang2023}.
These observations have been extended to other forms of electric and thermal transports~\cite{nakai2019nonreciprocal,zeng2019nonlinear,zeng2020,yu2019topological}.
Thus, the spontaneous Hall effect presents an important paradigm for topological response, where Berry-curved, non-centrosymmetric matter behaves in a manner that has no direct analog in Berry-flat matter under an applied magnetic field.
This effect has been experimentally studied in 
transition metal dichalcogenides~\cite{ma2019, kang2019, qin2021strain, tiwari2021giant, huang2020giant} and other topological materials~\cite{kiswandhi2021observation, he2021quantum, kumar2021room, ho2021hall}.

In the WKSM Ce$_3$Bi$_4$Pd$_3$, not only that the spontaneous Hall effect has been observed but its magnitude is extremely large~\cite{dzsaber2021giant}. 
Moreover, there is an $1\omega$ component that normally is associated with systems 
whose Berry curvature is not odd in ${\bf k}$, which arises when the 
time-reversal symmetry
is broken. Because the system is time-reversal invariant at the microscopic level, the effect cannot be understood within the existing theoretical framework~\cite{sodemann2015}.
An important 
feature of 
{the WKSM model} 
is that the Weyl nodes are pinned near the Fermi energy~\cite{lai2018weyl,grefe2020weyl}.
Thus, a small Fermi surface 
encloses the Weyl nodes, and
{the} Berry curvature density
{is high} on the Fermi surface. 
Moreover, for the generic case in the presence of a symmetry-allowed ``tilting potential'' \cite{dzsaber2021giant,grefe2020weyl},
the WKSM's small Fermi surface implies that
the system can
readily be driven beyond the perturbative regime 
{by} 
even 
the smallest 
applied electric fields. 


\begin{figure*}[!t]
\centering
\subfloat[]{%
\includegraphics[width=0.33\linewidth]{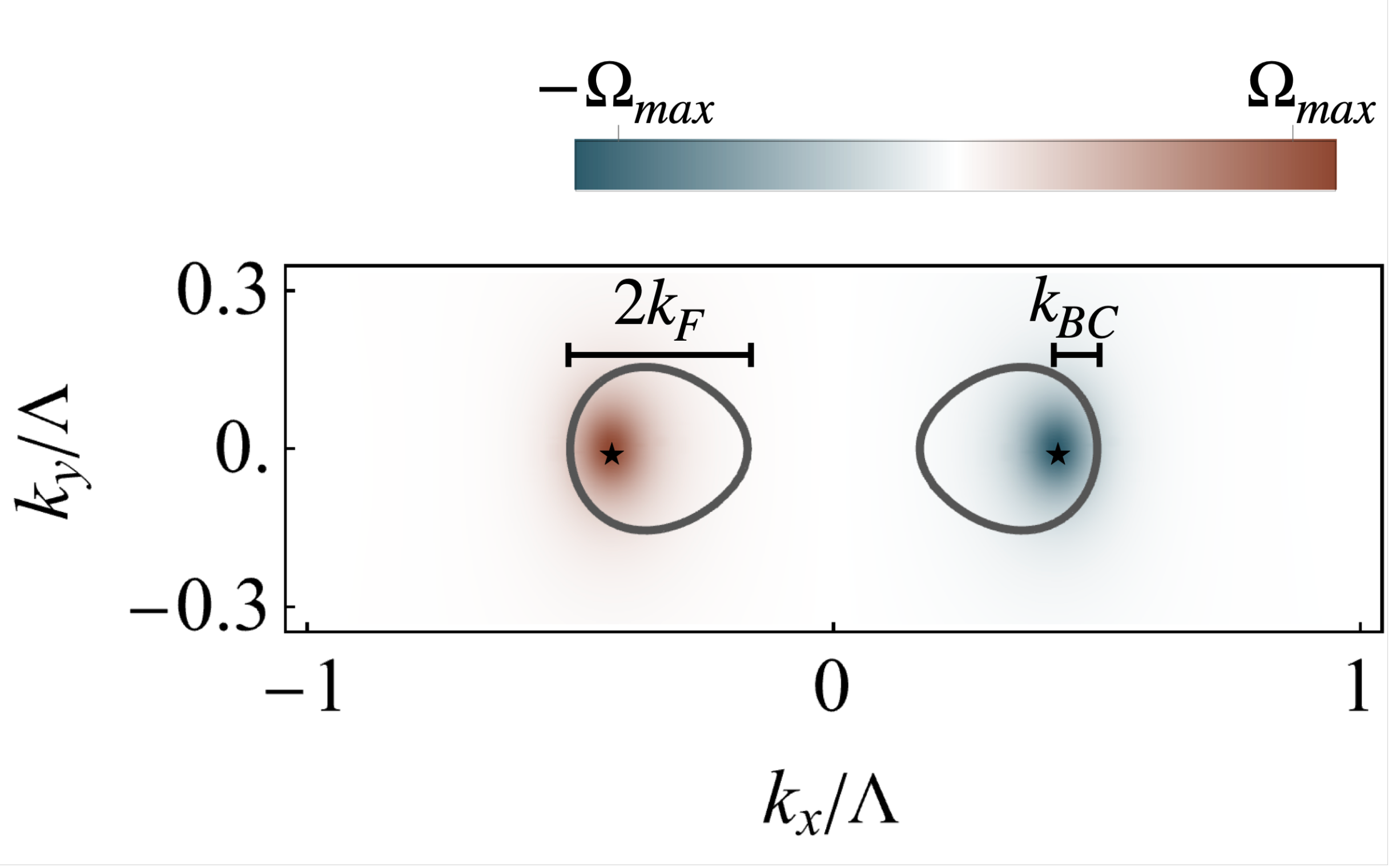}%
}
\hfill
\subfloat[]{%
\includegraphics[width=0.35\linewidth]{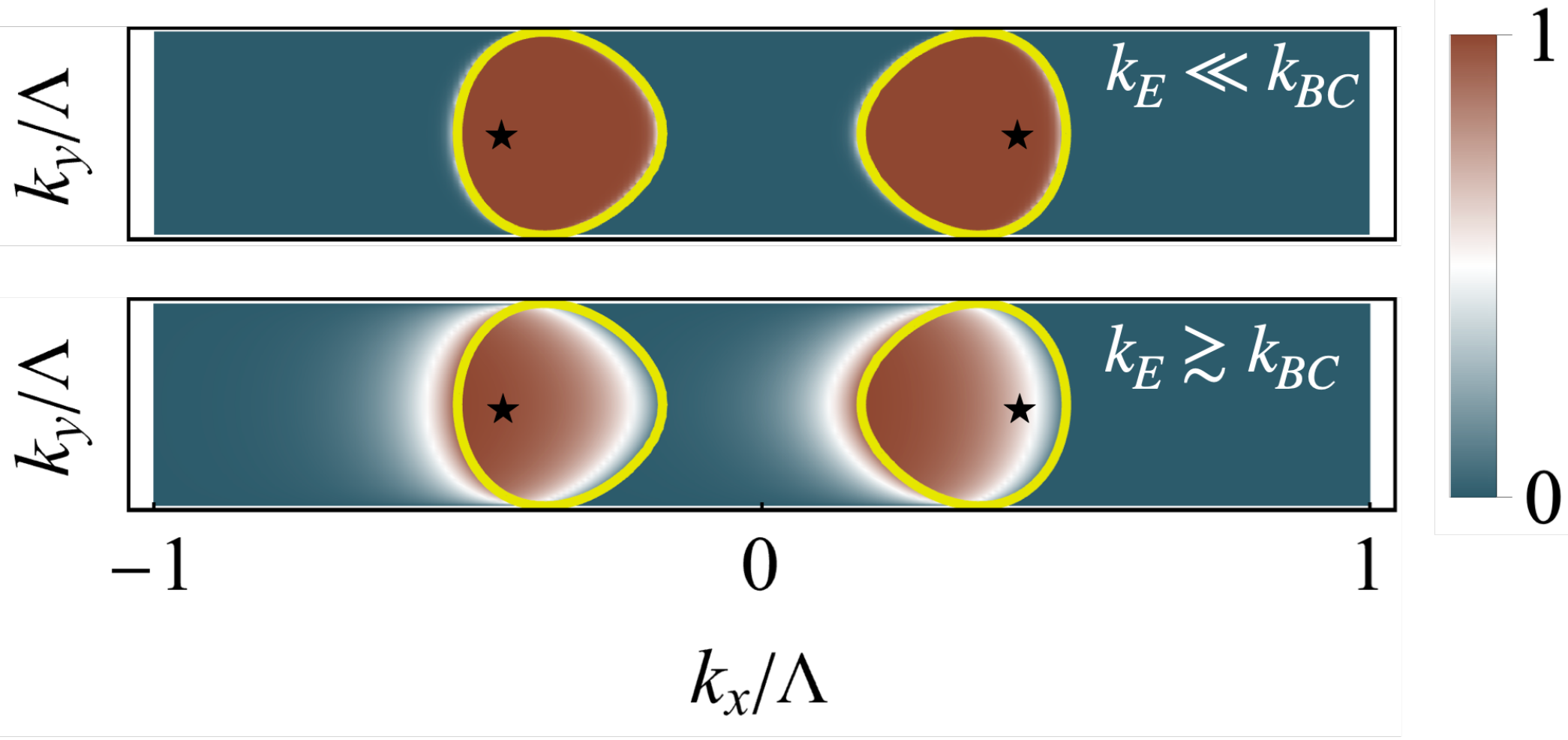}%
}
\hfill
\subfloat[]{%
\includegraphics[width=0.3\linewidth]{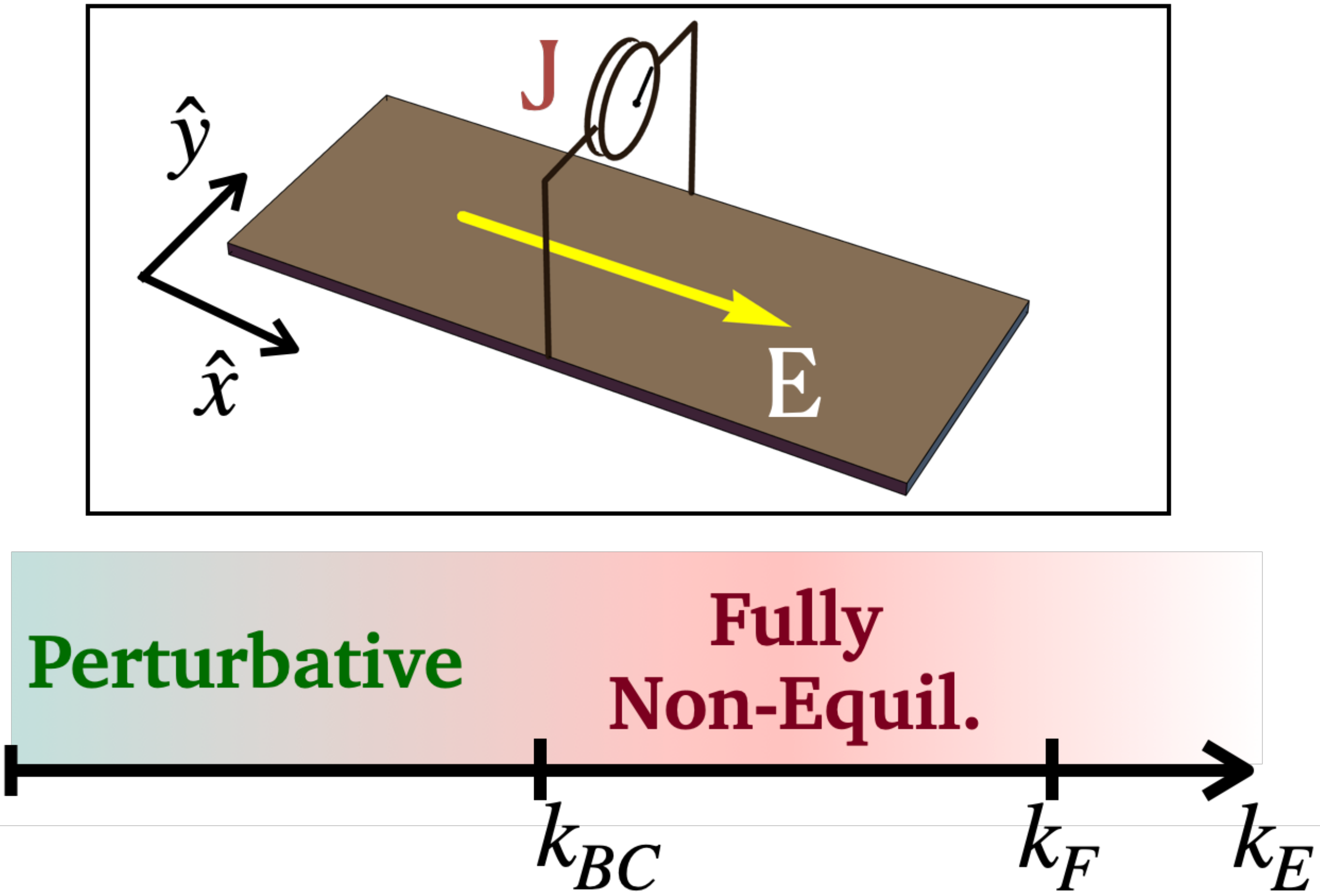}%
}
\caption{
Interplay between the applied $\bs E$-field and bandstructure of 
inversion-broken 
{semi}metals. 
(a) Position of the extrema of the Berry curvature density (star) relative to the Fermi surface (gray loops), which specifies the scale $k_\bc$. Here, we consider a filling that results in two Fermi pockets that are related by $k_x \mapsto -k_x$, $\Omega_{max}$ is the maximum  of the Berry curvature density, $\bs E = E \hat x$, and     $\Lambda$ is an ultraviolet momentum-cutoff. 
(b) Comparison of the $T=0$ electron distribution function at weak (top panel) vs. strong (bottom panel) fields. Here, the electric field is applied along $\hat x$ direction, and the (yellow) loops represent the Fermi pockets.
In (a) and (b) the momentum scales, $k_\bc$ and $k_F$, are expressed in units of $\Lambda$. 
(c) Schematic of the setup (inset) and distinct regimes of 
{topological} non-linear response as a function of the applied electric field ($k_E$). 
}
\label{fig:dipole-BC}
\end{figure*}

Thus motivated, here we advance a theory for a fully nonequilibrium (FNE) Hall response, which comes from the Berry curvature in the non-perturbative regime of the electric field.  
 {Here, positive and negative Berry curvature densities are no longer supported by the same number of current-carrying states.}
Specifically, we show that the 
spontaneous Hall current  crosses over to a quasi-linear scaling with $E$ over 
an extended regime of moderately strong fields.
In this FNE Hall response,  the relaxation time develops a spatial asymmetry which in turn generates an $1\omega$ response, even in the presence of microscopic TRS.
Our findings 
{identify a new type of}
non-linear topological response
{in TRS-preserving semimetallic systems}.

\para{\texorpdfstring{$\bs E$}{TEXT}-field response of inversion-broken
{semi}metals}
To illustrate the effect, we consider 
an electron-doped, 
multi-band system that preserves TRS  but breaks inversion symmetry due to  its non-centrosymmetric crystal structure.
We focus on 
a regime of 
a dilute concentration of electron carriers, 
in line with the motivation given in the introduction;
we will 
refer to this regime as a ``semimetal''.
For simplicity, we 
{consider} the two-dimensional (2D) case, though our 
central result also applies to three-dimensional systems.
A partially filled band is assumed to be well-isolated from the rest of the bands, such that the corresponding band-gap is the largest energy scale for the physics we discuss below. 
Although the net Berry curvature supported by this band would vanish due to the presence of TRS, the Berry curvature density is non-trivial in general, and it is responsible for the non-linear response to an applied 
$\bs E$ field.

We define the two key momentum scales that control 
{the spontaneous Hall} 
response, which are in addition to the Fermi momentum ($k_F$). 
The first scale is obtained by translating the electric field strength into a momentum scale, 
\begin{align}
k_E \coloneqq e E \tau_0/\hbar \, ,
\label{eq:kE}
\end{align} 
with $E = |\bs E|$ and $\tau_0$ ($e$) being the relaxation time in the absence of $\bs E$ (the quantum of electric charge). 
The second scale, $k_\bc$, is associated with the Berry curvature distribution, and it is defined as  the shortest distance between the extrema of the Berry curvature density and the Fermi surface along $\bs E$~[cf. Fig.~\ref{fig:dipole-BC}(a)].
{This scale is 
different from $k_F$ for 
the general case where a potential tilts the electronic band 
dispersion near its band bottom.}
These scales define {a}  ``weak-field regime'' as the limit $k_E \ll k_\bc$, where the electric field's effects can be treated perturbatively in terms of a small parameter $\epsilon \ll 1$, 
with
\begin{align}
\epsilon = k_E/k_\bc  \, .
\end{align}
In this regime, the effects of the driving force, i.e.  $\bs E$ in the present context, can be studied with respect to the global-equilibrium state.
By contrast, for $\epsilon \gtrsim 1$, the reference state is no longer at global equilibrium and the non-linear response enters a distinct regime, where 
non-linearities
arise due to a substantial change in the occupation of electronic modes supporting large Berry curvature densities.
Such non-linearities, 
{corresponding to the FNE responses that we focus on,}
have not been treated in prior theoretical studies.

Here, we use the Boltzmann equation within the relaxation time approximation to access both the perturbative 
{regime} in $\eps$
{and the FNE regime.}
Assuming $\bs E$ to be spatially uniform, the Boltzmann equation and the collision integral,
\begin{subequations}
\begin{align}
& \partial_t f(\bs k, t) + (\partial_t \bs k) \cdot \bs \nabla_k f(\bs k, t)  = \frac{f_0(\bs k) - f(\bs k, t)}{\tau(\bs k, t)} \, ,  \\
& \frac{1}{\tau(\bs k, t)} = \frac{2\pi}{\hbar[f(\bs k, t) - f_0(\bs k)]} \int \dd{\bs k'} \delta(\veps_k - \veps_{k'}) \nn \\
& \qquad \qquad \quad \times |U(\bs k - \bs k')|^2 \qty[f(\bs k, t) - f(\bs k', t)] \, ,
\end{align}
\label{eq:BE}
\end{subequations}
represent a set of coupled equations, whose general solution determines the non-perturbative $\bs E$-dependence of the {nonequilibrium} distribution function, $f$, and the relaxation time, $\tau$ 
(we have suppressed explicit reference to the $\bs E$-dependence of $f$ and $\tau$ for notational convenience). 
We note that the collision term in the Boltzmann equation arises from elastic scatterings with  impurities -- $U(\bs k)$ is the impurity potential -- and $f_0(\bs k)$ is  the 
{equilibrium} Fermi-Dirac distribution function.
At very weak fields, for an isotropic $U(\bs k)$, the relaxation time is effectively a constant for momentum states in the vicinity of the unperturbed Fermi surface.
At stronger fields, however, the $\bs E$-driven distortion of $f$ influences the scattering rate, even for an isotropic $U(\bs k)$, and $\tau$
develops a non-trivial dependence on the Fermi surface coordinate(s), as demonstrated in the Supplementary Materials (SM)~\cite{sm}.

\para{DC spontaneous Hall response}
The qualitatively important role played by $\eps$ in non-linear response functions is most readily accessible in the DC limit.
It is convenient to introduce $g(\bs k, \bs E) = f(\bs k,  \bs E) - f_0(\bs k)$, which allows us to re-write the Boltzmann equation as
\begin{align}
- \frac{e}{\hbar} \bs E \cdot \bs \nabla_k g(\bs k,  \bs E) + \frac{g(\bs k,  \bs E)}{\tau(\bs k, \bs E)}  = \frac{e}{\hbar} \bs E \cdot \bs \nabla_k f_0(\bs k) \, ,
\label{eq:g-Boltzmann}
\end{align}
where we have utilized the semiclassical relationship, $\partial_t \bs k = - e \bs E/\hbar$.
Note that we do not assume $g$ to be small compared to $f_0$. 
Since \eq{eq:g-Boltzmann} is insensitive to $k_\bc$, a perturbative solution to $g$  can be organized in terms of the parameter  $\tilde k_E \coloneqq k_E/\Lambda$ (henceforth, we will denote dimensionless ratios `$X/\Lambda$' as `$\tilde X$'), where $\Lambda$ is the ultraviolet momentum-cutoff associated with, for example, the Brillouin zone size.
Hypothesizing a solution of the form
$g(\bs k, \bs E) = \sum_{n=1}^\infty \tilde k_E^n ~g_n(\bs k)$, we obtain $g_n = \partial_{ \tilde k_x}^n f_0$.
Here, in the interest of a perturbative solution, we have set $\tau(\bs k, \bs E) = \tau_0$
and, without loss of generality, assumed $\bs E = \hat x E$.
Thus, the transverse current takes the form 
\begin{align}
J_y(\bs E) =  \frac{e \Lambda}{\tau_0} \sum_{n=1}^\infty (-1)^n \tilde k_E^{n+1} \int \dd{\bs k} f_0(\bs k) \partial_{ \tilde k_x}^n \Om_{xy}(\bs k).
\label{eq:DC-Jy-perturb}
\end{align}
The leading order non-vanishing term  in the expansion dominates over the rest in the weak-field limit, as shown in the SM~\cite{sm}.

In order to understand the behavior of $J_y$ beyond the weak-field limit, we directly solve the differential equation in \eq{eq:g-Boltzmann},
with
$\tau(\bs k, \bs E)$ still taken as 
$\tau_0$,
to obtain 
$g =  e^{\tilde k_x/\tilde k_E} \int_{\tilde k_x}^1 \dd{\tilde q} e^{-\tilde q/\tilde k_E} \partial_{\tilde q} f_0$.
{The nonequilibrium distribution function is}  depicted in Fig.~\ref{fig:dipole-BC}(b).
While both 
longitudinal and transverse response functions would be expected to exhibit a crossover behavior as a function of $k_E/k_F$, 
{the} transverse  response functions that originate from the  Berry curvature density are {additionally} impacted by 
$k_\bc$ {because of the $\bs E$-field driven occupation and depletion of states supporting high Berry curvature densities (see SM~\cite{sm}). }

{Importantly,} if the $\bs E$-driven  reorganization of the Fermi sea significantly affects momentum states supporting high Berry curvature density, then an additional crossover behavior 
is expected as a function of $\eps$.
Here, we will consider instances where $k_\bc \ll k_F$ such that the sources of crossover behaviors are well separated along the $k_E$ axis, 
{and the regime of Berry-curvature-enabled FNE response sets in at  $k_E$ well below $k_F$,} 
as summarized in Fig.~\ref{fig:dipole-BC}(c).

\para{AC spontaneous Hall response} 
Since both $k_\bc$ and $k_F$ continue to operate when the DC electric field is replaced by an AC field, the qualitative aspects of response functions discussed in  the DC limit remains valid for an oscillating field, $\bs E(t) = \hat x E \cos{\omega t}$. 
Previous works have addressed AC conductivity in non-centrosymmetric systems through a perturbative expansion of the $n\omega$ harmonics of $g$  in terms of $k_E$~\cite{sodemann2015,nakai2019nonreciprocal,zhang2023}.
A similar strategy in the
{FNE} regime fails, however, since the coupling among the $n\omega$ harmonics is controlled by $k_E$, and a self-consistent truncation in $n$ is not possible~\cite{sm}.
Here, we take a distinct approach and solve for the time-dependent distribution function directly from the Boltzmann equation, and eventually decompose the transverse current into its $n\om$-harmonics to obtain the spontaneous AC Hall current. 

We utilize the  time-dependent form of $\bs k$, $\bs k(t) = \bs k_0 -   \hat k_x (k_E/\tau_0  \omega) \sin{\om t}$ with $\bs k_0 \coloneqq \bs k(0)$, to  express $f$ as an explicitly time-dependent function, and solve the linear differential equation, 
$df(\bs k_0, t)/dt = [f_0(\bs k_0 - \bs \phi_t) - f(\bs k_0, t)]/{\tau(t)}$, 
to obtain 
\begin{align}
f(\bs k_0, t) = A_{-1}(t) \qty[
f_0(\bs k_0) + \int_0^t \dd{t'} f_0(\bs k_0- \bs \phi_{t'})\frac{A_1(t') }{\tau(t')}
] \, ,
\end{align}
where reference to $\bs E$-dependence is suppressed for notational convenience, $\bs \phi_t \coloneqq \hat k_x (k_E/\tau_0  \omega) \sin{\om t}$, and we have defined $A_s(t) = \exp{s\int_0^t \frac{\dd{t'}}{\tau(t')} }$.
{The $n$-th harmonic of the transverse current is given by} 
$\bar J_y^{(n)} = \int_{t_0 - \pi/\om}^{t_0 + \pi/\om} e^{-i n \om t} J_y(t)$, where $\om t_0, t_0/\tau_0  \gg 1$ such that the system enters a (time-periodic) steady state, and the transverse current is $J_y( t) = (e^2 E/\hbar) \cos{\om t} \allowbreak \int \dd{\bs k_0} f(\bs k_0 + \bs \phi_t, t) \Om_{xy}(\bs k_0)$.

\para{Odd-\texorpdfstring{$\om$}{TEXT} vs. even-\texorpdfstring{$\om$}{TEXT} response}
\begin{figure}[!t]
\centering
\subfloat[\label{fig:DC}]{%
\includegraphics[width=0.9\columnwidth]{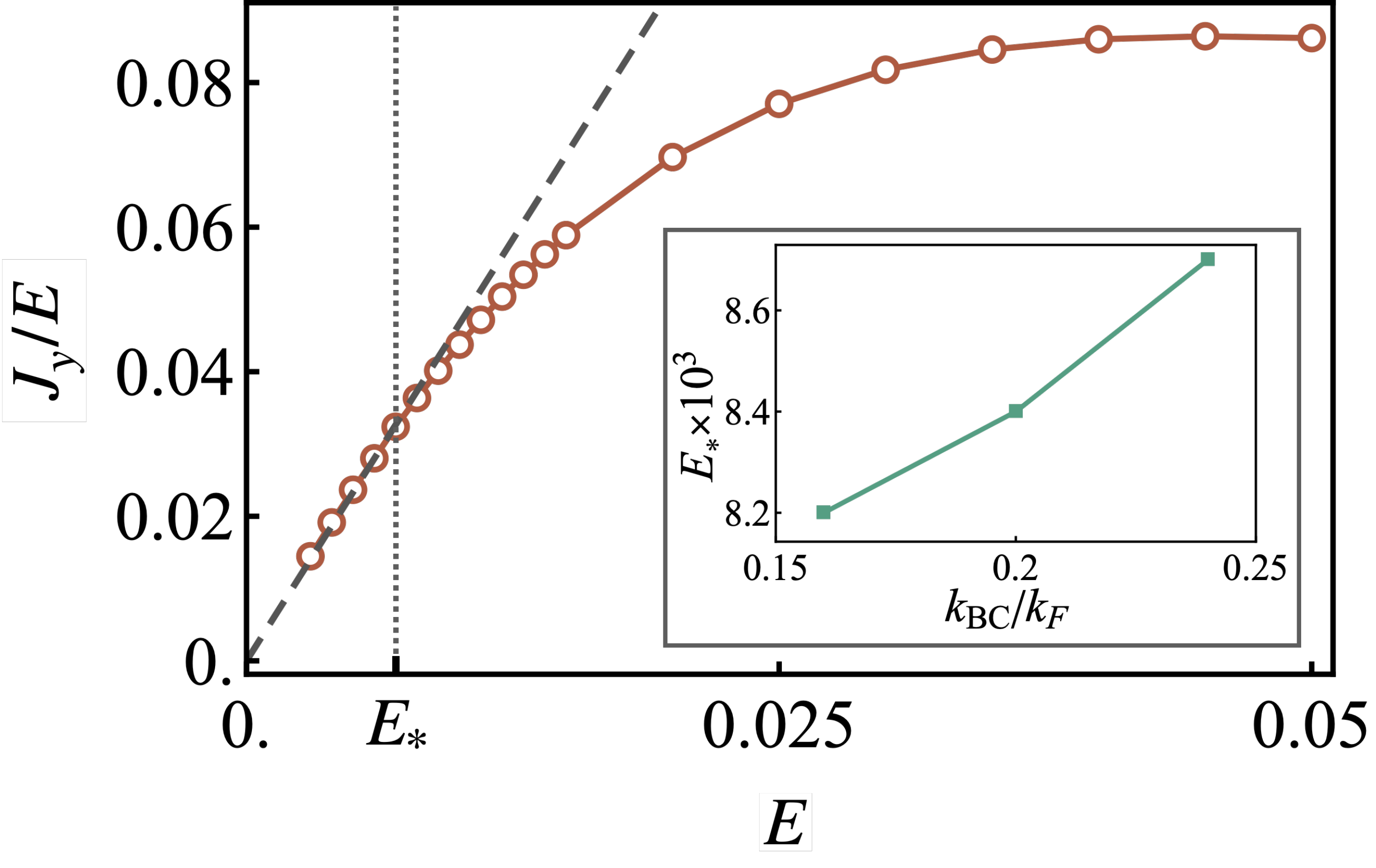}%
}
\hfill
\subfloat[\label{fig:quadrupole}]{%
\includegraphics[width=0.85\columnwidth]{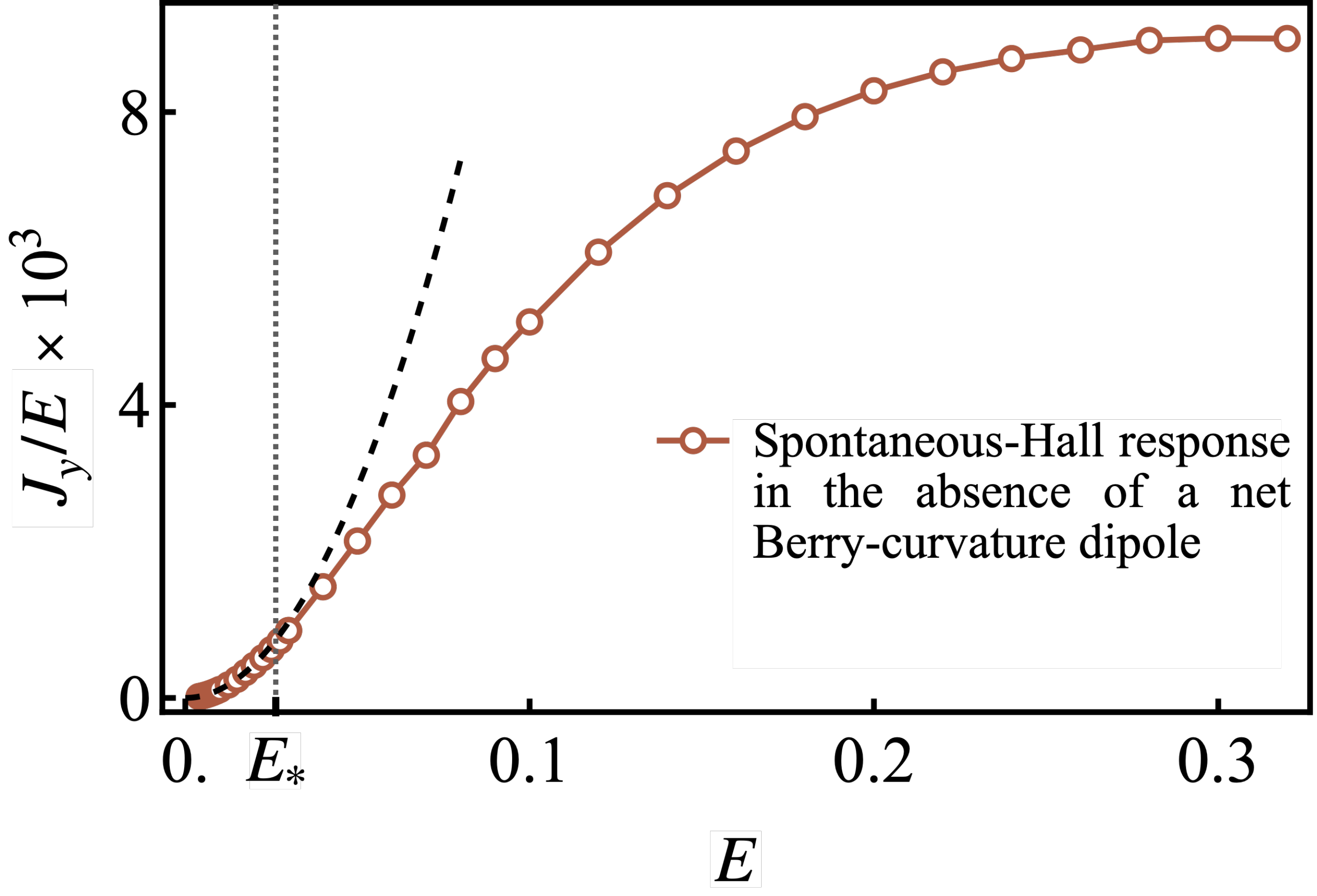}%
}
\caption{DC spontaneous Hall conductivity ($J_y/E$) as a function of the magnitude of the applied electric field, $E$. 
(a) In generic inversion-broken metals, while at weak-fields $J_y/E \propto E$, at stronger fields it tends scale as $E^a$ with $a \approx 0$. 
The dashed line marks the trend of $J_y/E$ if its low-field scaling were to continue to hold at stronger fields.
The departure from the weak-field response is controlled by a crossover scale, $E_*$, which, in turn, depends on the proximity of the extrema of the  Berry curvature density to the Fermi surface, $k_{\text{BC}}$ [see inset; $k_\bc$ is tuned at a fixed $k_F$ by tuning both $\mu$ and $B_0/B_x$ in Eq.~\eqref{eq:h0}]. 
Here, $J_y/E$ is measured in units of $e^2/\hbar$,  $E$ and $E_*$ are measured in units of $\hbar/(e\tau_0)$ [cf. Eq.~\eqref{eq:kE}], and the calculations are performed at $T=0$.  
(b) $J_y/E$ in an $\mc{C}_4 \mc{T}$-symmetric system where a Berry quadrupole controls the leading response at weak fields ($J_y/E \sim E^2$; dashed curve). 
At stronger fields, $J_y/E$ behaves similarly as in (a). 
Here, $J_y/E$ [same units as in (a)] is scaled by a factor of $10^3$.
}
\label{fig:DC-Jy}
\end{figure}

 {In general, the relaxation time is $\bs k$-dependent and sensitive to the symmetries of the system. Moreover, under an applied electric field, it}
develops a dependence on $\bs k \cdot \hat E$, even if $\tau$ is assumed to be  spatially uniform at equilibrium~\cite{sm}.
This directional dependence  {at a finite $\bs E$} is a result of nonequilibrium steady-state conditions imposed by the field, where scattering rates can differ for states at $\mathbf{k}$ and -$\mathbf{k}$.
We show that the spatial non-uniformity of $\tau$  {generated by $\bs E$}
leads to a non-trivial odd-$\om$ response~\footnote{In the limit where the relaxation time is a constant, $\tau(t) = \tau_*$, we observe that $J_y(t + \pi/\om) = J_y(t) + e^{-t/\tau_*} I(t)$.
Thus, the contribution of $J_y(t)$ to $\bar J_y^{(2n+1)}$ is exponentially suppressed at $t \gg \tau_* \sim \tau_0$.
Therefore, for a spatiotemporally uniform relaxation time only even harmonics of $J_y$ are appreciable  deep in the steady state.
It is easily checked that this non-perturbative outcome is consistent with the perturbative results obtained in the limit of a uniform $\tau$~\cite{nakai2019nonreciprocal,zhang2023}.
}. 
The fact that a finite  odd-$\omega$ response is generated by the $\bs E$-dependent corrections to the relaxation time indicates that  the reference state for this response is not in {global} equilibrium.

 {In the interest of universality, we ignore specific space-group symmetry allowed $\bs k$-dependence of $\tau$, and} find that a minimal  {$\bs E$-driven} modification of the relaxation time takes the form
\begin{align}
1/\tau(\bs k) = 1/\tau_0 + \alpha(\bs E) ~\mbox{sign}(\hat E \cdot \bs k),
\label{eq:tau}
\end{align} 
with  $\lim_{E\to 0} \alpha(\bs E) = 0$~\cite{sm}.
By making the $t$-dependence of $\bs k$ and $\bs E$ explicit, it can be  checked that  $J_y^{an}$ is no longer asymptotically periodic in $\pi/\om$, {in contrast to the limit where $\tau = \tau_0$}.
Consequently, $1\omega$ and higher odd-$\om$ responses 
become {nonzero.}
Since {this nonzero result is} dependent on $\alpha \neq 0$, 
the response in  odd-$\om$ channels is proportional to $|\alpha(\bs E)|^b$ with $b>0$.
Therefore, at weak fields, $\bar J_y^{(2n+1)}$ scales with a higher exponent of $E$ than anticipated by previous perturbative calculations~\cite{sm}.
We note that 
the $1\om$ response, resulting from a lack of specific parity of the collision integral under $\bs k \to - \bs k$ in the presence of the $\bs E$-field, may be interpreted as a consequence of irreversibility in a dissipitative system.
This implies a monotonic change of entropy, and the emergence of an ``arrow'' of time at the thermodynamic scale, even though the  microscopic time-reversal symmetry is preserved~\cite{dzsaber2021giant}.

\para{Specific model calculations}
In order to explicate the general results above, we consider a two-band $k.p$ model of an 2D system without inversion symmetry, 
{as specified by the following Hamiltonian:}
\begin{align}
H(\bs k) =&   \qty(M_x - B_x k_x^2 ) \sig_1 + A k_y \sig_2 + \qty(M_y - B_y k_y^2 ) \sig_3 \nn \\
& ~ + B_0 k_x^2 \sigma_0 \,.
\label{eq:h0}
\end{align}
Here, $\qty{A, M_j, B_j}$ are momentum-independent parameters, and $\sigma_j$ ($\sigma_0$) is the $j$-th Pauli ($2\times 2$ identity) matrix.
Since $H^*(- \bs k) = H(\bs k)$,  
the Hamiltonian preserves a spinless time-reversal symmetry (TRS) with $\mc T^2 = +1$. 
Consequently, $\Om_{xy}(-\bs k) = - \Om_{xy}(\bs k)$, which prevents a 
{nonzero} anomalous Hall response.
$H(\bs k)$  breaks the 2D inversion  symmetry, however, and, thus, will  support a non-linear Hall response.

In the vicinity of the topological critical point {at $M_y = 0$}, the Berry curvature is finite, but remains strongly peaked near the minima (maxima) of the conduction (valence) band.
Here, we will consider all parameters in the Hamiltonian to be positive, and introduce a chemical potential, $\mu$, that crosses the conduction band. 
$k_\bc$ and $k_F$ are most prominently controlled by the dimensioless ratios  $M_y/(B_x \Lam^2)$ and  $B_0/B_x$, and  $\mu$.
We select a parameter window which results in $k_\bc \ll k_F$, such that the extrema of the Berry curvature density lie close to the Fermi surface, as shown in Fig.~\ref{fig:dipole-BC}(a).

In Fig.~\ref{fig:DC} we plot the spontaneous Hall current $J_y$ as a function of $E$ in the DC limit.
The non-vanishing contribution from the Berry curvature dipole, $\partial_{k_x} \Om_{xy}$, leads to a 
{nonzero} spontaneous Hall response at weak fields such that  $J_y  \sim E^2$.
As $k_E$ approaches $k_\bc$ from below, the occupation number in the vicinity of the extrema of the Berry curvature density is strongly modified [c.f.  Fig.~\ref{fig:dipole-BC}(b)], which leads to a deviation from the weak-field scaling behavior, as anticipated 
by the analysis of the perturbative expansion in \eq{eq:DC-Jy-perturb} {(see the SM~\cite{sm} for details)}.
Eventually, at even stronger fields where $k_E \sim k_F$,  $J_y$ scales quasi-linearly with $E$.
{The response of the system for $k_E \gtrsim k_\bc$, the net response must be understood relative to a nonequilibrium steady state. Thus, we obtain an FNE Hall response. In this regime, the response  can no longer be attributed to the dipole of the Berry curvature. Instead, the contributions from all moments of the Berry curvature are  comparable.
}

\begin{figure}[!t]
\centering
\subfloat[]{%
\includegraphics[width=0.7\columnwidth]{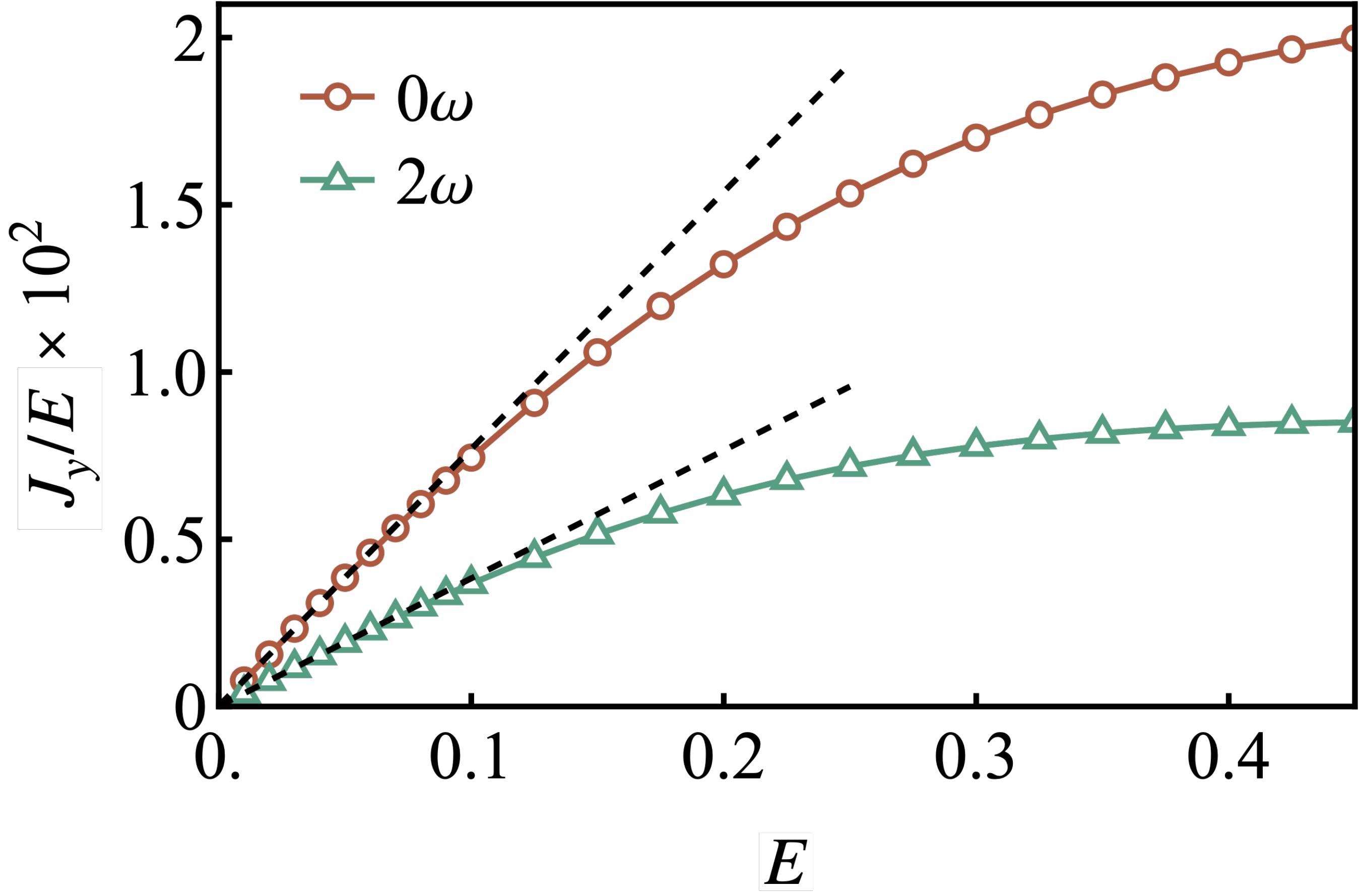}%
}
\hfill
\subfloat[]{%
\includegraphics[width=0.7\columnwidth]{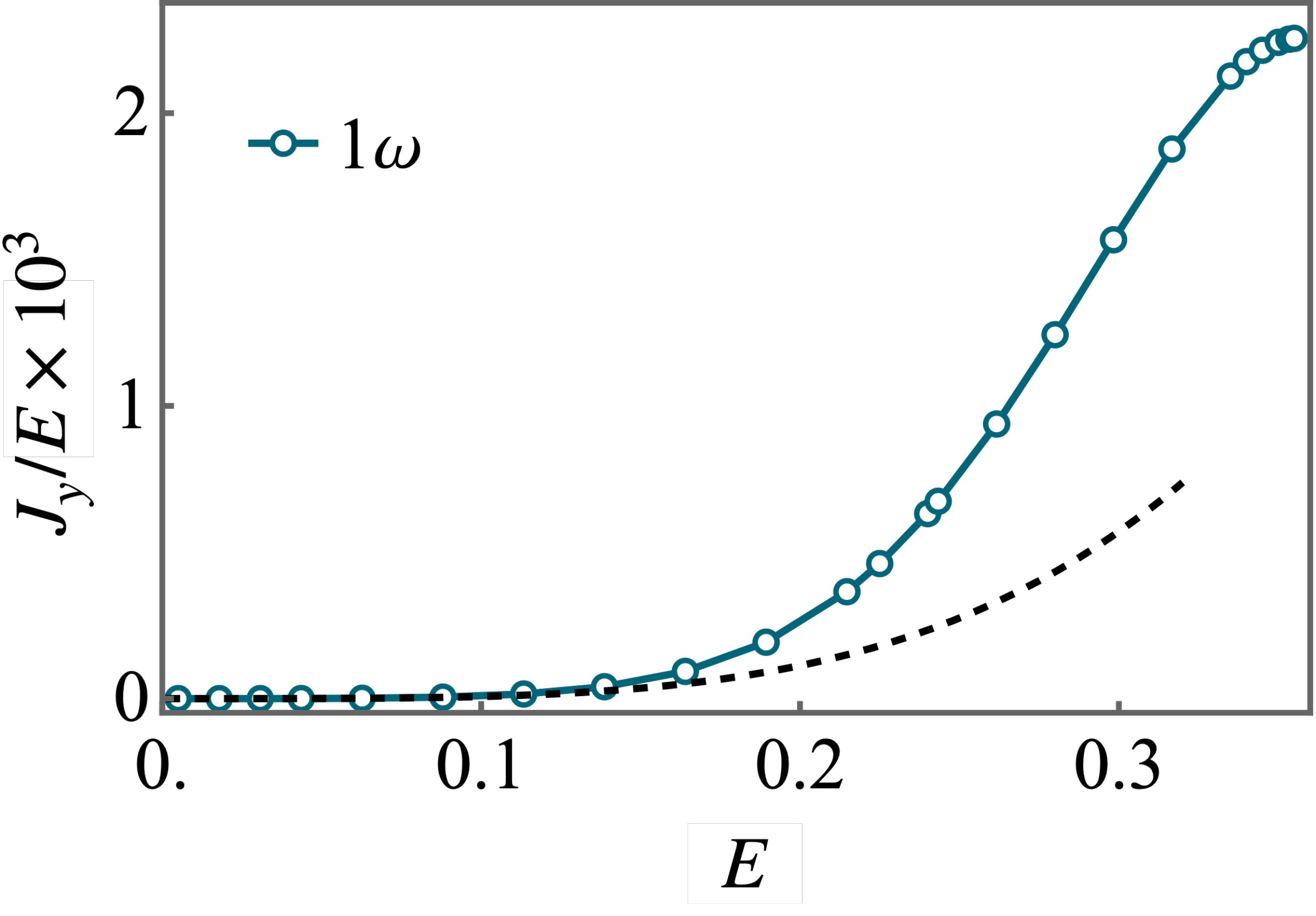}
}
\hfill
\caption{AC spontaneous Hall  conductivity as a function of the strength of the applied field, $E$.
(a) For weak fields both the $0\om$ and $2\om$ components scale as $E^1$ (dashed lines), as anticipated from the perturbative calculation.
At strong fields corresponding to $k_E \gtrsim k_F$ the conductivity becomes weakly $E$-dependent, with the response resembling the Hall  response in a TRS broken system.
(b) $1\om$ response for a inhomogeneous relaxation time.
In this regime  $\bar J_y/E \sim E^4$ (dashed curve), which depends on our choice of $\alpha(\bs E) = \alpha_0 E^2$ [cf. Eq.~\eqref{eq:tau}]. 
Here, the units are identical to that in Fig.~\ref{fig:DC-Jy}, and calculations are performed at $T>0$. 
}
\label{fig:AC-Hall}
\end{figure}

For an AC electric field and 
{equilibrium} relaxation time only the even-$\om$ channels exhibit  non-trivial responses in the steady state. 
The corresponding $0\om$ ($m=0$) and the $2\om$ ($m=1$) responses are demonstrated in Fig.~\ref{fig:AC-Hall}(a).
At weak fields, both $\bar J_y^{(0)}$ and $\bar J_y^{(2)}$ are proportional to $E^2$, in agreement with the perturbative results~\cite{sodemann2015}.
At stronger fields, the 
non-perturbatively obtained  $n\omega$ channels exhibit a qualitatively similar $E$-dependence as in the DC limit.

For {an $\bs E$-field driven} spatially non-uniform $\tau$, both even- and odd-$\om$ responses are non-trivial. 
{Based on the explicit solution of the 1D case,}
{we choose $\alpha(\bs E) = \alpha_0 \qty(E \cos{\om t})^2$ with  $\alpha_0 > 0$ being a model-dependent constant.}
While the even-$\om$ channels undergo a quantitative change in their behavior, spontaneous Hall 
responses in  the odd-$\om$ channels become finite and $\bar J_y^{(1)} \sim E^5$ for $\eps \ll 1$, as shown in Fig.~\ref{fig:AC-Hall}(b) {and discussed further in the SM~\cite{sm}.}
{We note that the precise weak-field scaling of $\bar J_y^{(1)}$ with $E$ is determined by the form of non-uniformity present in the relaxation time. 

{Crucially,}
irrespective of the weak-field scaling of the Hall current, the intermediate to high-field response is independent of the form of $\alpha(\bs E)$. 
Thus, there is a sense of universality in the FNE response where the spontaneous  Hall current, in both the DC and AC limits, tend to scale quasi-linearly at sufficiently strong fields. 
}

\para{Response from higher Berry-multipoles}
The breakdown of the perturbatively obtained scaling of $J_y$ is, in fact, a general feature of spontaneous Hall response, and it is present in metals with additional point-group symmetries that prevent the Berry curvature dipole from generating a non-trivial response at weak fields.
In Fig.~\ref{fig:quadrupole}, we exemplify this behavior by computing the  spontaneous DC Hall current for a system with $\mc{C}_4 \mc{T}$ symmetry (for details of the model, see section V of the SI~\cite{sm}) that leads to a Berry curvature quadrupole controlling the leading order response at weak-fields, and $J_y \sim E^3$~\cite{sm}. 
At stronger fields the perturbative scaling no longer holds, and the response becomes qualitatively similar to that obtained in the absence of the $\mc{C}_4 \mc{T}$ symmetry (cf. Fig.~\ref{fig:DC}).
{Thus, the universality of the FNE response identified above is expected to extend  across all non-centrosymmetric space groups that support non-trivial spontaneous Hall response.}

\para{Discussion}
Our formalism for the FNE Hall response can be directly extended to TRS-preserving non-centrosymmetric three dimensional (3D) metals,  obtained by doping a gapped bandstructure, and we expect to find qualitatively similar behaviors in suitable parameter regimes.
The 2D model we studied may be considered as a representative of the 2D planes  that separate the Weyl points in a non-centrosymmetric Weyl semimetal (WSM) {as explicated in the SM~\cite{sm}.
Therefore, the spontaneous Hall  response of the 3D semimetal may be viewed as the sum of the corresponding response of all such 2D metallic planes.}
Thus, our results are strongly indicative of the physics behind very similar observations in the Weyl-Kondo semimetal Ce$_3$Bi$_4$Pd$_3$~\cite{dzsaber2021giant}.
A direct analysis of such WKSMs, however, requires a 
systematic account of multiband effects, because the peak of the Berry curvature density corresponds to the band-crossing points, and the 
role of Kondo physics.
 {We note that a comprehensive account of multiband effects  may  introduce additional sources of non-linearity through the $\bs E$-field dependent changes to the quantum geometry. 
Similarly, correlation effects beyond those that can be absorbed into the collision integral and the renormalizations to bare parameters may require a more systematic treatment~\cite{morimoto2018}.
Our work sets the stage for future studies on both refinements,
which will be important for explorations of nonlinear topological response in the FNE regime in more realistic settings.}


 {Strong-field effects on quantum transport is also relevant to moir\'{e}-materials due to the greatly reduced size of the Brillouin zone~\cite{phong2023quantum}. 
Although recent works have explored the role of Bloch oscillations in such systems~\cite{de2024floquet}, the analogs of the effects studied here remain unexplored. 
Therefore, non-centrosymmetric moir\'{e} materials provides a highly tunable and experimentally accessible platform for realizing FNE. 
Finally, the details of the space-group symmetry of various materials platforms would generally lead of a symmetry-allowed $\bs k$-dependence of the relaxation time. 
A detailed exploration of such applications and generalizations is left to future works.
}

\para{Conclusion}
To summarize, we 
have theoretically demonstrated 
a fully nonequilibrium Hall response from Berry curvature in
{non-centrosymmetric and TRS-preserving} {semi}metals.
We elucidated the key role played by two intrinsic momentum scales   $k_\bc$ and $k_F$,
with $k_\bc$ arising from the distribution of the Berry curvature and  setting the scale beyond which  
the system {enters an FNE  regime}. 
In this 
{FNE} regime, the reference state for the Hall response is itself out of equilibrium. 
Such a non-equilibrium reference state leads to a spatial asymmetry in the relaxation time, which in turn generates a 
{nonzero} $1\omega$ response, even in the presence of microscopic TRS.
Our work provides the foundation to understand the nonequilibrium
transport properties of Weyl-Kondo semimetals. Because strong correlations naturally produce emergent electronic excitations near the Fermi energies, we can expect that the conditions for the fully nonequilibrium transport responses advanced here are naturally met in topological semimetals driven by strong correlations. As such, we expect our work to have broader implications for metallic topological systems in strongly correlated settings.

\acknowledgments
We thank
 Yuan Fang, Haoyu Hu, and Elio K\"onig  for helpful discussions.
 Work at Rice has primarily been supported by the Air Force Office of Scientific Research under Grant No.
FA9550-21-1-0356 (conceptualization and model construction, S.S.,\,L.C.,\,Y.W.,\,C.S. and Q.S.),
by the National Science Foundation
under Grant No. DMR-2220603 (model calculations, S.S. and L.C.),
by the Robert A. Welch Foundation Grant No. C-1411 (Q.S.) and the Vannevar Bush Faculty Fellowship ONR-VB N00014-23-1-2870 (Q.S.). The
majority of the computational calculations have been performed on the Shared University Grid
at Rice funded by NSF under Grant EIA-0216467, a partnership between Rice University, Sun
Microsystems, and Sigma Solutions, Inc., the Big-Data Private-Cloud Research Cyberinfrastructure
MRI-award funded by NSF under Grant No. CNS-1338099, and the Extreme Science and
Engineering Discovery Environment (XSEDE) by NSF under Grant No. DMR170109. 
S.P. acknowledges funding by the European Union (ERC, CorMeTop, project 101055088), the Austrian Science Fund (FWF) through the project I5868-N-FOR 5249 (QUAST), and the Air Force Office of Scientific Research under Grant No.\ FA8655-24-1-7018 (CorTopS).
C.S. acknowledges support from Iowa State University  and Ames National Laboratory start-up funding. 
All authors acknowledge 
the hospitality of the Kavli Institute for Theoretical Physics, UCSB,
supported in part
by the National Science Foundation under Grant No. NSF PHY-1748958,
 during the program ``A Quantum Universe in
a Crystal: Symmetry and Topology across the Correlation Spectrum."
S.S. and Q.S. also 
acknowledge the hospitality of the Aspen Center for Physics, which is supported by the National Science Foundation under Grant No. PHY-2210452.
{The schematic of the setup in Fig.~\ref{fig:dipole-BC}(c) is adapted from Ref.~\cite{figure}.}


\bibliography{refer.bib}

\begin{thebibliography}{39}%
\makeatletter
\providecommand \@ifxundefined [1]{%
 \@ifx{#1\undefined}
}%
\providecommand \@ifnum [1]{%
 \ifnum #1\expandafter \@firstoftwo
 \else \expandafter \@secondoftwo
 \fi
}%
\providecommand \@ifx [1]{%
 \ifx #1\expandafter \@firstoftwo
 \else \expandafter \@secondoftwo
 \fi
}%
\providecommand \natexlab [1]{#1}%
\providecommand \enquote  [1]{``#1''}%
\providecommand \bibnamefont  [1]{#1}%
\providecommand \bibfnamefont [1]{#1}%
\providecommand \citenamefont [1]{#1}%
\providecommand \href@noop [0]{\@secondoftwo}%
\providecommand \href [0]{\begingroup \@sanitize@url \@href}%
\providecommand \@href[1]{\@@startlink{#1}\@@href}%
\providecommand \@@href[1]{\endgroup#1\@@endlink}%
\providecommand \@sanitize@url [0]{\catcode `\\12\catcode `\$12\catcode
  `\&12\catcode `\#12\catcode `\^12\catcode `\_12\catcode `\%12\relax}%
\providecommand \@@startlink[1]{}%
\providecommand \@@endlink[0]{}%
\providecommand \url  [0]{\begingroup\@sanitize@url \@url }%
\providecommand \@url [1]{\endgroup\@href {#1}{\urlprefix }}%
\providecommand \urlprefix  [0]{URL }%
\providecommand \Eprint [0]{\href }%
\providecommand \doibase [0]{http://dx.doi.org/}%
\providecommand \selectlanguage [0]{\@gobble}%
\providecommand \bibinfo  [0]{\@secondoftwo}%
\providecommand \bibfield  [0]{\@secondoftwo}%
\providecommand \translation [1]{[#1]}%
\providecommand \BibitemOpen [0]{}%
\providecommand \bibitemStop [0]{}%
\providecommand \bibitemNoStop [0]{.\EOS\space}%
\providecommand \EOS [0]{\spacefactor3000\relax}%
\providecommand \BibitemShut  [1]{\csname bibitem#1\endcsname}%
\let\auto@bib@innerbib\@empty
\bibitem [{\citenamefont {Paschen}\ and\ \citenamefont {Si}(2021)}]{Pas21.1}%
  \BibitemOpen
  \bibfield  {author} {\bibinfo {author} {\bibfnamefont {S.}~\bibnamefont
  {Paschen}}\ and\ \bibinfo {author} {\bibfnamefont {Q.}~\bibnamefont {Si}},\
  }\href {\doibase 10.1038/s42254-020-00262-6} {\bibfield  {journal} {\bibinfo
  {journal} {Nat.\ Rev.\ Phys.}\ }\textbf {\bibinfo {volume} {3}},\ \bibinfo
  {pages} {9} (\bibinfo {year} {2021})}\BibitemShut {NoStop}%
\bibitem [{\citenamefont {Witczak-Krempa}\ \emph {et~al.}(2014)\citenamefont
  {Witczak-Krempa}, \citenamefont {Chen}, \citenamefont {Kim},\ and\
  \citenamefont {Balents}}]{Wit14.1}%
  \BibitemOpen
  \bibfield  {author} {\bibinfo {author} {\bibfnamefont {W.}~\bibnamefont
  {Witczak-Krempa}}, \bibinfo {author} {\bibfnamefont {G.}~\bibnamefont
  {Chen}}, \bibinfo {author} {\bibfnamefont {Y.~B.}\ \bibnamefont {Kim}}, \
  and\ \bibinfo {author} {\bibfnamefont {L.}~\bibnamefont {Balents}},\ }\href
  {\doibase 10.1146/annurev-conmatphys-020911-125138} {\bibfield  {journal}
  {\bibinfo  {journal} {{Annu.\ Rev.\ Condens.\ Matter Phys.}}\ }\textbf
  {\bibinfo {volume} {5}},\ \bibinfo {pages} {57} (\bibinfo {year}
  {2014})}\BibitemShut {NoStop}%
\bibitem [{\citenamefont {Lai}\ \emph {et~al.}(2018)\citenamefont {Lai},
  \citenamefont {Grefe}, \citenamefont {Paschen},\ and\ \citenamefont
  {Si}}]{lai2018weyl}%
  \BibitemOpen
  \bibfield  {author} {\bibinfo {author} {\bibfnamefont {H.-H.}\ \bibnamefont
  {Lai}}, \bibinfo {author} {\bibfnamefont {S.~E.}\ \bibnamefont {Grefe}},
  \bibinfo {author} {\bibfnamefont {S.}~\bibnamefont {Paschen}}, \ and\
  \bibinfo {author} {\bibfnamefont {Q.}~\bibnamefont {Si}},\ }\href {\doibase
  10.1073/pnas.1715851115} {\bibfield  {journal} {\bibinfo  {journal}
  {Proceedings of the National Academy of Sciences}\ }\textbf {\bibinfo
  {volume} {115}},\ \bibinfo {pages} {93} (\bibinfo {year} {2018})},\ \Eprint
  {http://arxiv.org/abs/https://www.pnas.org/doi/pdf/10.1073/pnas.1715851115}
  {https://www.pnas.org/doi/pdf/10.1073/pnas.1715851115} \BibitemShut {NoStop}%
\bibitem [{\citenamefont {Chen}\ \emph {et~al.}(2022)\citenamefont {Chen},
  \citenamefont {Setty}, \citenamefont {Hu}, \citenamefont {Vergniory},
  \citenamefont {Grefe}, \citenamefont {Fischer}, \citenamefont {Yan},
  \citenamefont {Eguchi}, \citenamefont {Prokofiev}, \citenamefont {Paschen},
  \citenamefont {Cano},\ and\ \citenamefont {Si}}]{chen2022topological}%
  \BibitemOpen
  \bibfield  {author} {\bibinfo {author} {\bibfnamefont {L.}~\bibnamefont
  {Chen}}, \bibinfo {author} {\bibfnamefont {C.}~\bibnamefont {Setty}},
  \bibinfo {author} {\bibfnamefont {H.}~\bibnamefont {Hu}}, \bibinfo {author}
  {\bibfnamefont {M.~G.}\ \bibnamefont {Vergniory}}, \bibinfo {author}
  {\bibfnamefont {S.~E.}\ \bibnamefont {Grefe}}, \bibinfo {author}
  {\bibfnamefont {L.}~\bibnamefont {Fischer}}, \bibinfo {author} {\bibfnamefont
  {X.}~\bibnamefont {Yan}}, \bibinfo {author} {\bibfnamefont {G.}~\bibnamefont
  {Eguchi}}, \bibinfo {author} {\bibfnamefont {A.}~\bibnamefont {Prokofiev}},
  \bibinfo {author} {\bibfnamefont {S.}~\bibnamefont {Paschen}}, \bibinfo
  {author} {\bibfnamefont {J.}~\bibnamefont {Cano}}, \ and\ \bibinfo {author}
  {\bibfnamefont {Q.}~\bibnamefont {Si}},\ }\href {\doibase
  10.1038/s41567-022-01743-4} {\bibfield  {journal} {\bibinfo  {journal}
  {Nature Physics}\ }\textbf {\bibinfo {volume} {18}},\ \bibinfo {pages} {1341}
  (\bibinfo {year} {2022})}\BibitemShut {NoStop}%
\bibitem [{\citenamefont {Dzsaber}\ \emph {et~al.}(2017)\citenamefont
  {Dzsaber}, \citenamefont {Prochaska}, \citenamefont {Sidorenko},
  \citenamefont {Eguchi}, \citenamefont {Svagera}, \citenamefont {Waas},
  \citenamefont {Prokofiev}, \citenamefont {Si},\ and\ \citenamefont
  {Paschen}}]{dzsaber2017kondo}%
  \BibitemOpen
  \bibfield  {author} {\bibinfo {author} {\bibfnamefont {S.}~\bibnamefont
  {Dzsaber}}, \bibinfo {author} {\bibfnamefont {L.}~\bibnamefont {Prochaska}},
  \bibinfo {author} {\bibfnamefont {A.}~\bibnamefont {Sidorenko}}, \bibinfo
  {author} {\bibfnamefont {G.}~\bibnamefont {Eguchi}}, \bibinfo {author}
  {\bibfnamefont {R.}~\bibnamefont {Svagera}}, \bibinfo {author} {\bibfnamefont
  {M.}~\bibnamefont {Waas}}, \bibinfo {author} {\bibfnamefont {A.}~\bibnamefont
  {Prokofiev}}, \bibinfo {author} {\bibfnamefont {Q.}~\bibnamefont {Si}}, \
  and\ \bibinfo {author} {\bibfnamefont {S.}~\bibnamefont {Paschen}},\ }\href
  {\doibase 10.1103/PhysRevLett.118.246601} {\bibfield  {journal} {\bibinfo
  {journal} {Phys. Rev. Lett.}\ }\textbf {\bibinfo {volume} {118}},\ \bibinfo
  {pages} {246601} (\bibinfo {year} {2017})}\BibitemShut {NoStop}%
\bibitem [{\citenamefont {Dzsaber}\ \emph {et~al.}(2021)\citenamefont
  {Dzsaber}, \citenamefont {Yan}, \citenamefont {Taupin}, \citenamefont
  {Eguchi}, \citenamefont {Prokofiev}, \citenamefont {Shiroka}, \citenamefont
  {Blaha}, \citenamefont {Rubel}, \citenamefont {Grefe}, \citenamefont {Lai},
  \citenamefont {Si},\ and\ \citenamefont {Paschen}}]{dzsaber2021giant}%
  \BibitemOpen
  \bibfield  {author} {\bibinfo {author} {\bibfnamefont {S.}~\bibnamefont
  {Dzsaber}}, \bibinfo {author} {\bibfnamefont {X.}~\bibnamefont {Yan}},
  \bibinfo {author} {\bibfnamefont {M.}~\bibnamefont {Taupin}}, \bibinfo
  {author} {\bibfnamefont {G.}~\bibnamefont {Eguchi}}, \bibinfo {author}
  {\bibfnamefont {A.}~\bibnamefont {Prokofiev}}, \bibinfo {author}
  {\bibfnamefont {T.}~\bibnamefont {Shiroka}}, \bibinfo {author} {\bibfnamefont
  {P.}~\bibnamefont {Blaha}}, \bibinfo {author} {\bibfnamefont
  {O.}~\bibnamefont {Rubel}}, \bibinfo {author} {\bibfnamefont {S.~E.}\
  \bibnamefont {Grefe}}, \bibinfo {author} {\bibfnamefont {H.-H.}\ \bibnamefont
  {Lai}}, \bibinfo {author} {\bibfnamefont {Q.}~\bibnamefont {Si}}, \ and\
  \bibinfo {author} {\bibfnamefont {S.}~\bibnamefont {Paschen}},\ }\href
  {\doibase 10.1073/pnas.2013386118} {\bibfield  {journal} {\bibinfo  {journal}
  {Proceedings of the National Academy of Sciences}\ }\textbf {\bibinfo
  {volume} {118}},\ \bibinfo {pages} {e2013386118} (\bibinfo {year} {2021})},\
  \Eprint
  {http://arxiv.org/abs/https://www.pnas.org/doi/pdf/10.1073/pnas.2013386118}
  {https://www.pnas.org/doi/pdf/10.1073/pnas.2013386118} \BibitemShut {NoStop}%
\bibitem [{\citenamefont {Grefe}\ \emph {et~al.}(2020)\citenamefont {Grefe},
  \citenamefont {Lai}, \citenamefont {Paschen},\ and\ \citenamefont
  {Si}}]{grefe2020weyl}%
  \BibitemOpen
  \bibfield  {author} {\bibinfo {author} {\bibfnamefont {S.~E.}\ \bibnamefont
  {Grefe}}, \bibinfo {author} {\bibfnamefont {H.-H.}\ \bibnamefont {Lai}},
  \bibinfo {author} {\bibfnamefont {S.}~\bibnamefont {Paschen}}, \ and\
  \bibinfo {author} {\bibfnamefont {Q.}~\bibnamefont {Si}},\ }\href {\doibase
  10.1103/PhysRevB.101.075138} {\bibfield  {journal} {\bibinfo  {journal}
  {Phys. Rev. B}\ }\textbf {\bibinfo {volume} {101}},\ \bibinfo {pages}
  {075138} (\bibinfo {year} {2020})}\BibitemShut {NoStop}%
\bibitem [{\citenamefont {Jungwirth}\ \emph {et~al.}(2002)\citenamefont
  {Jungwirth}, \citenamefont {Niu},\ and\ \citenamefont
  {MacDonald}}]{jungwirth2002anomalous}%
  \BibitemOpen
  \bibfield  {author} {\bibinfo {author} {\bibfnamefont {T.}~\bibnamefont
  {Jungwirth}}, \bibinfo {author} {\bibfnamefont {Q.}~\bibnamefont {Niu}}, \
  and\ \bibinfo {author} {\bibfnamefont {A.}~\bibnamefont {MacDonald}},\ }\href
  {\doibase 10.1103/PhysRevLett.88.207208} {\bibfield  {journal} {\bibinfo
  {journal} {Physical review letters}\ }\textbf {\bibinfo {volume} {88}},\
  \bibinfo {pages} {207208} (\bibinfo {year} {2002})}\BibitemShut {NoStop}%
\bibitem [{\citenamefont {Fang}\ \emph {et~al.}(2003)\citenamefont {Fang},
  \citenamefont {Nagaosa}, \citenamefont {Takahashi}, \citenamefont {Asamitsu},
  \citenamefont {Mathieu}, \citenamefont {Ogasawara}, \citenamefont {Yamada},
  \citenamefont {Kawasaki}, \citenamefont {Tokura},\ and\ \citenamefont
  {Terakura}}]{fang2003anomalous}%
  \BibitemOpen
  \bibfield  {author} {\bibinfo {author} {\bibfnamefont {Z.}~\bibnamefont
  {Fang}}, \bibinfo {author} {\bibfnamefont {N.}~\bibnamefont {Nagaosa}},
  \bibinfo {author} {\bibfnamefont {K.~S.}\ \bibnamefont {Takahashi}}, \bibinfo
  {author} {\bibfnamefont {A.}~\bibnamefont {Asamitsu}}, \bibinfo {author}
  {\bibfnamefont {R.}~\bibnamefont {Mathieu}}, \bibinfo {author} {\bibfnamefont
  {T.}~\bibnamefont {Ogasawara}}, \bibinfo {author} {\bibfnamefont
  {H.}~\bibnamefont {Yamada}}, \bibinfo {author} {\bibfnamefont
  {M.}~\bibnamefont {Kawasaki}}, \bibinfo {author} {\bibfnamefont
  {Y.}~\bibnamefont {Tokura}}, \ and\ \bibinfo {author} {\bibfnamefont
  {K.}~\bibnamefont {Terakura}},\ }\href {\doibase 10.1126/science.1089408}
  {\bibfield  {journal} {\bibinfo  {journal} {Science}\ }\textbf {\bibinfo
  {volume} {302}},\ \bibinfo {pages} {92} (\bibinfo {year} {2003})}\BibitemShut
  {NoStop}%
\bibitem [{\citenamefont {Yao}\ \emph {et~al.}(2004)\citenamefont {Yao},
  \citenamefont {Kleinman}, \citenamefont {MacDonald}, \citenamefont {Sinova},
  \citenamefont {Jungwirth}, \citenamefont {Wang}, \citenamefont {Wang},\ and\
  \citenamefont {Niu}}]{yao2004first}%
  \BibitemOpen
  \bibfield  {author} {\bibinfo {author} {\bibfnamefont {Y.}~\bibnamefont
  {Yao}}, \bibinfo {author} {\bibfnamefont {L.}~\bibnamefont {Kleinman}},
  \bibinfo {author} {\bibfnamefont {A.}~\bibnamefont {MacDonald}}, \bibinfo
  {author} {\bibfnamefont {J.}~\bibnamefont {Sinova}}, \bibinfo {author}
  {\bibfnamefont {T.}~\bibnamefont {Jungwirth}}, \bibinfo {author}
  {\bibfnamefont {D.-s.}\ \bibnamefont {Wang}}, \bibinfo {author}
  {\bibfnamefont {E.}~\bibnamefont {Wang}}, \ and\ \bibinfo {author}
  {\bibfnamefont {Q.}~\bibnamefont {Niu}},\ }\href {\doibase
  10.1103/PhysRevLett.92.037204} {\bibfield  {journal} {\bibinfo  {journal}
  {Physical review letters}\ }\textbf {\bibinfo {volume} {92}},\ \bibinfo
  {pages} {037204} (\bibinfo {year} {2004})}\BibitemShut {NoStop}%
\bibitem [{\citenamefont {Thonhauser}\ \emph {et~al.}(2005)\citenamefont
  {Thonhauser}, \citenamefont {Ceresoli}, \citenamefont {Vanderbilt},\ and\
  \citenamefont {Resta}}]{thonhauser2005orbital}%
  \BibitemOpen
  \bibfield  {author} {\bibinfo {author} {\bibfnamefont {T.}~\bibnamefont
  {Thonhauser}}, \bibinfo {author} {\bibfnamefont {D.}~\bibnamefont
  {Ceresoli}}, \bibinfo {author} {\bibfnamefont {D.}~\bibnamefont
  {Vanderbilt}}, \ and\ \bibinfo {author} {\bibfnamefont {R.}~\bibnamefont
  {Resta}},\ }\href {\doibase 10.1103/PhysRevLett.95.137205} {\bibfield
  {journal} {\bibinfo  {journal} {Physical review letters}\ }\textbf {\bibinfo
  {volume} {95}},\ \bibinfo {pages} {137205} (\bibinfo {year}
  {2005})}\BibitemShut {NoStop}%
\bibitem [{\citenamefont {Xiao}\ \emph {et~al.}(2006)\citenamefont {Xiao},
  \citenamefont {Yao}, \citenamefont {Fang},\ and\ \citenamefont
  {Niu}}]{xiao2006berry}%
  \BibitemOpen
  \bibfield  {author} {\bibinfo {author} {\bibfnamefont {D.}~\bibnamefont
  {Xiao}}, \bibinfo {author} {\bibfnamefont {Y.}~\bibnamefont {Yao}}, \bibinfo
  {author} {\bibfnamefont {Z.}~\bibnamefont {Fang}}, \ and\ \bibinfo {author}
  {\bibfnamefont {Q.}~\bibnamefont {Niu}},\ }\href {\doibase
  10.1103/PhysRevLett.97.026603} {\bibfield  {journal} {\bibinfo  {journal}
  {Physical review letters}\ }\textbf {\bibinfo {volume} {97}},\ \bibinfo
  {pages} {026603} (\bibinfo {year} {2006})}\BibitemShut {NoStop}%
\bibitem [{\citenamefont {Xiao}\ \emph {et~al.}(2010)\citenamefont {Xiao},
  \citenamefont {Chang},\ and\ \citenamefont {Niu}}]{xiao2010}%
  \BibitemOpen
  \bibfield  {author} {\bibinfo {author} {\bibfnamefont {D.}~\bibnamefont
  {Xiao}}, \bibinfo {author} {\bibfnamefont {M.~C.}\ \bibnamefont {Chang}}, \
  and\ \bibinfo {author} {\bibfnamefont {Q.}~\bibnamefont {Niu}},\ }\href
  {\doibase 10.1103/RevModPhys.82.1959} {\bibfield  {journal} {\bibinfo
  {journal} {Rev. Mod. Phys.}\ }\textbf {\bibinfo {volume} {82}},\ \bibinfo
  {pages} {1959} (\bibinfo {year} {2010})}\BibitemShut {NoStop}%
\bibitem [{\citenamefont {Nagaosa}\ \emph {et~al.}(2010)\citenamefont
  {Nagaosa}, \citenamefont {Sinova}, \citenamefont {Onoda}, \citenamefont
  {MacDonald},\ and\ \citenamefont {Ong}}]{nagaosa2010anomalous}%
  \BibitemOpen
  \bibfield  {author} {\bibinfo {author} {\bibfnamefont {N.}~\bibnamefont
  {Nagaosa}}, \bibinfo {author} {\bibfnamefont {J.}~\bibnamefont {Sinova}},
  \bibinfo {author} {\bibfnamefont {S.}~\bibnamefont {Onoda}}, \bibinfo
  {author} {\bibfnamefont {A.~H.}\ \bibnamefont {MacDonald}}, \ and\ \bibinfo
  {author} {\bibfnamefont {N.~P.}\ \bibnamefont {Ong}},\ }\href {\doibase
  10.1103/RevModPhys.82.1539} {\bibfield  {journal} {\bibinfo  {journal}
  {Reviews of modern physics}\ }\textbf {\bibinfo {volume} {82}},\ \bibinfo
  {pages} {1539} (\bibinfo {year} {2010})}\BibitemShut {NoStop}%
\bibitem [{\citenamefont {Deyo}\ \emph {et~al.}(2009)\citenamefont {Deyo},
  \citenamefont {Golub}, \citenamefont {Ivchenko},\ and\ \citenamefont
  {Spivak}}]{deyo2009}%
  \BibitemOpen
  \bibfield  {author} {\bibinfo {author} {\bibfnamefont {E.}~\bibnamefont
  {Deyo}}, \bibinfo {author} {\bibfnamefont {L.}~\bibnamefont {Golub}},
  \bibinfo {author} {\bibfnamefont {E.}~\bibnamefont {Ivchenko}}, \ and\
  \bibinfo {author} {\bibfnamefont {B.}~\bibnamefont {Spivak}},\ }\href
  {\doibase 10.48550/arXiv.0904.1917} {\bibfield  {journal} {\bibinfo
  {journal} {arXiv preprint arXiv:0904.1917}\ } (\bibinfo {year} {2009}),\
  10.48550/arXiv.0904.1917}\BibitemShut {NoStop}%
\bibitem [{\citenamefont {Sodemann}\ and\ \citenamefont
  {Fu}(2015)}]{sodemann2015}%
  \BibitemOpen
  \bibfield  {author} {\bibinfo {author} {\bibfnamefont {I.}~\bibnamefont
  {Sodemann}}\ and\ \bibinfo {author} {\bibfnamefont {L.}~\bibnamefont {Fu}},\
  }\href {\doibase 10.1103/PhysRevLett.115.216806} {\bibfield  {journal}
  {\bibinfo  {journal} {Physical review letters}\ }\textbf {\bibinfo {volume}
  {115}},\ \bibinfo {pages} {216806} (\bibinfo {year} {2015})}\BibitemShut
  {NoStop}%
\bibitem [{\citenamefont {Low}\ \emph {et~al.}(2015)\citenamefont {Low},
  \citenamefont {Jiang},\ and\ \citenamefont {Guinea}}]{low2015}%
  \BibitemOpen
  \bibfield  {author} {\bibinfo {author} {\bibfnamefont {T.}~\bibnamefont
  {Low}}, \bibinfo {author} {\bibfnamefont {Y.}~\bibnamefont {Jiang}}, \ and\
  \bibinfo {author} {\bibfnamefont {F.}~\bibnamefont {Guinea}},\ }\href
  {\doibase 10.1103/PhysRevB.92.235447} {\bibfield  {journal} {\bibinfo
  {journal} {Physical Review B}\ }\textbf {\bibinfo {volume} {92}},\ \bibinfo
  {pages} {235447} (\bibinfo {year} {2015})}\BibitemShut {NoStop}%
\bibitem [{\citenamefont {Zhang}\ \emph {et~al.}(2023)\citenamefont {Zhang},
  \citenamefont {Gao}, \citenamefont {Xie}, \citenamefont {Po},\ and\
  \citenamefont {Law}}]{zhang2023}%
  \BibitemOpen
  \bibfield  {author} {\bibinfo {author} {\bibfnamefont {C.-P.}\ \bibnamefont
  {Zhang}}, \bibinfo {author} {\bibfnamefont {X.-J.}\ \bibnamefont {Gao}},
  \bibinfo {author} {\bibfnamefont {Y.-M.}\ \bibnamefont {Xie}}, \bibinfo
  {author} {\bibfnamefont {H.~C.}\ \bibnamefont {Po}}, \ and\ \bibinfo {author}
  {\bibfnamefont {K.~T.}\ \bibnamefont {Law}},\ }\href {\doibase
  10.48550/arXiv.2012.15628} {\bibfield  {journal} {\bibinfo  {journal}
  {Physical Review B}\ }\textbf {\bibinfo {volume} {107}},\ \bibinfo {pages}
  {115142} (\bibinfo {year} {2023})}\BibitemShut {NoStop}%
\bibitem [{\citenamefont {Nakai}\ and\ \citenamefont
  {Nagaosa}(2019)}]{nakai2019nonreciprocal}%
  \BibitemOpen
  \bibfield  {author} {\bibinfo {author} {\bibfnamefont {R.}~\bibnamefont
  {Nakai}}\ and\ \bibinfo {author} {\bibfnamefont {N.}~\bibnamefont
  {Nagaosa}},\ }\href {\doibase 10.1103/PhysRevB.99.115201} {\bibfield
  {journal} {\bibinfo  {journal} {Physical Review B}\ }\textbf {\bibinfo
  {volume} {99}},\ \bibinfo {pages} {115201} (\bibinfo {year}
  {2019})}\BibitemShut {NoStop}%
\bibitem [{\citenamefont {Zeng}\ \emph {et~al.}(2019)\citenamefont {Zeng},
  \citenamefont {Nandy}, \citenamefont {Taraphder},\ and\ \citenamefont
  {Tewari}}]{zeng2019nonlinear}%
  \BibitemOpen
  \bibfield  {author} {\bibinfo {author} {\bibfnamefont {C.}~\bibnamefont
  {Zeng}}, \bibinfo {author} {\bibfnamefont {S.}~\bibnamefont {Nandy}},
  \bibinfo {author} {\bibfnamefont {A.}~\bibnamefont {Taraphder}}, \ and\
  \bibinfo {author} {\bibfnamefont {S.}~\bibnamefont {Tewari}},\ }\href
  {\doibase 10.1103/PhysRevB.100.245102} {\bibfield  {journal} {\bibinfo
  {journal} {Physical Review B}\ }\textbf {\bibinfo {volume} {100}},\ \bibinfo
  {pages} {245102} (\bibinfo {year} {2019})}\BibitemShut {NoStop}%
\bibitem [{\citenamefont {Zeng}\ \emph {et~al.}(2020)\citenamefont {Zeng},
  \citenamefont {Nandy},\ and\ \citenamefont {Tewari}}]{zeng2020}%
  \BibitemOpen
  \bibfield  {author} {\bibinfo {author} {\bibfnamefont {C.}~\bibnamefont
  {Zeng}}, \bibinfo {author} {\bibfnamefont {S.}~\bibnamefont {Nandy}}, \ and\
  \bibinfo {author} {\bibfnamefont {S.}~\bibnamefont {Tewari}},\ }\href
  {\doibase 10.1103/PhysRevResearch.2.032066} {\bibfield  {journal} {\bibinfo
  {journal} {Physical Review Research}\ }\textbf {\bibinfo {volume} {2}},\
  \bibinfo {pages} {032066} (\bibinfo {year} {2020})}\BibitemShut {NoStop}%
\bibitem [{\citenamefont {Yu}\ \emph {et~al.}(2019)\citenamefont {Yu},
  \citenamefont {Zhu}, \citenamefont {You}, \citenamefont {Low},\ and\
  \citenamefont {Su}}]{yu2019topological}%
  \BibitemOpen
  \bibfield  {author} {\bibinfo {author} {\bibfnamefont {X.-Q.}\ \bibnamefont
  {Yu}}, \bibinfo {author} {\bibfnamefont {Z.-G.}\ \bibnamefont {Zhu}},
  \bibinfo {author} {\bibfnamefont {J.-S.}\ \bibnamefont {You}}, \bibinfo
  {author} {\bibfnamefont {T.}~\bibnamefont {Low}}, \ and\ \bibinfo {author}
  {\bibfnamefont {G.}~\bibnamefont {Su}},\ }\href {\doibase
  10.1103/PhysRevB.99.201410} {\bibfield  {journal} {\bibinfo  {journal}
  {Physical Review B}\ }\textbf {\bibinfo {volume} {99}},\ \bibinfo {pages}
  {201410} (\bibinfo {year} {2019})}\BibitemShut {NoStop}%
\bibitem [{\citenamefont {Ma}\ \emph {et~al.}(2019)\citenamefont {Ma},
  \citenamefont {Xu}, \citenamefont {Shen}, \citenamefont {MacNeill},
  \citenamefont {Fatemi}, \citenamefont {Chang}, \citenamefont {Mier~Valdivia},
  \citenamefont {Wu}, \citenamefont {Du}, \citenamefont {Hsu} \emph
  {et~al.}}]{ma2019}%
  \BibitemOpen
  \bibfield  {author} {\bibinfo {author} {\bibfnamefont {Q.}~\bibnamefont
  {Ma}}, \bibinfo {author} {\bibfnamefont {S.-Y.}\ \bibnamefont {Xu}}, \bibinfo
  {author} {\bibfnamefont {H.}~\bibnamefont {Shen}}, \bibinfo {author}
  {\bibfnamefont {D.}~\bibnamefont {MacNeill}}, \bibinfo {author}
  {\bibfnamefont {V.}~\bibnamefont {Fatemi}}, \bibinfo {author} {\bibfnamefont
  {T.-R.}\ \bibnamefont {Chang}}, \bibinfo {author} {\bibfnamefont {A.~M.}\
  \bibnamefont {Mier~Valdivia}}, \bibinfo {author} {\bibfnamefont
  {S.}~\bibnamefont {Wu}}, \bibinfo {author} {\bibfnamefont {Z.}~\bibnamefont
  {Du}}, \bibinfo {author} {\bibfnamefont {C.-H.}\ \bibnamefont {Hsu}},  \emph
  {et~al.},\ }\href {\doibase 10.1038/s41586-018-0807-6} {\bibfield  {journal}
  {\bibinfo  {journal} {Nature}\ }\textbf {\bibinfo {volume} {565}},\ \bibinfo
  {pages} {337} (\bibinfo {year} {2019})}\BibitemShut {NoStop}%
\bibitem [{\citenamefont {Kang}\ \emph {et~al.}(2019)\citenamefont {Kang},
  \citenamefont {Li}, \citenamefont {Sohn}, \citenamefont {Shan},\ and\
  \citenamefont {Mak}}]{kang2019}%
  \BibitemOpen
  \bibfield  {author} {\bibinfo {author} {\bibfnamefont {K.}~\bibnamefont
  {Kang}}, \bibinfo {author} {\bibfnamefont {T.}~\bibnamefont {Li}}, \bibinfo
  {author} {\bibfnamefont {E.}~\bibnamefont {Sohn}}, \bibinfo {author}
  {\bibfnamefont {J.}~\bibnamefont {Shan}}, \ and\ \bibinfo {author}
  {\bibfnamefont {K.~F.}\ \bibnamefont {Mak}},\ }\href {\doibase
  10.1038/s41563-019-0294-7} {\bibfield  {journal} {\bibinfo  {journal} {Nature
  materials}\ }\textbf {\bibinfo {volume} {18}},\ \bibinfo {pages} {324}
  (\bibinfo {year} {2019})}\BibitemShut {NoStop}%
\bibitem [{\citenamefont {Qin}\ \emph {et~al.}(2021)\citenamefont {Qin} \emph
  {et~al.}}]{qin2021strain}%
  \BibitemOpen
  \bibfield  {author} {\bibinfo {author} {\bibfnamefont {M.-S.}\ \bibnamefont
  {Qin}} \emph {et~al.},\ }\href@noop {} {\bibfield  {journal} {\bibinfo
  {journal} {Chin. Phys. Lett.}\ }\textbf {\bibinfo {volume} {38}},\ \bibinfo
  {pages} {017301} (\bibinfo {year} {2021})}\BibitemShut {NoStop}%
\bibitem [{\citenamefont {Tiwari}\ \emph {et~al.}(2021)\citenamefont {Tiwari}
  \emph {et~al.}}]{tiwari2021giant}%
  \BibitemOpen
  \bibfield  {author} {\bibinfo {author} {\bibfnamefont {A.}~\bibnamefont
  {Tiwari}} \emph {et~al.},\ }\href@noop {} {\bibfield  {journal} {\bibinfo
  {journal} {Nat. Commun.}\ }\textbf {\bibinfo {volume} {12}},\ \bibinfo
  {pages} {2049} (\bibinfo {year} {2021})}\BibitemShut {NoStop}%
\bibitem [{\citenamefont {Huang}\ \emph {et~al.}(2020)\citenamefont {Huang}
  \emph {et~al.}}]{huang2020giant}%
  \BibitemOpen
  \bibfield  {author} {\bibinfo {author} {\bibfnamefont {M.}~\bibnamefont
  {Huang}} \emph {et~al.},\ }\href@noop {} {\  (\bibinfo {year} {2020})},\
  \Eprint {http://arxiv.org/abs/2006.05615} {arXiv:2006.05615
  [cond-mat.mtrl-sci]} \BibitemShut {NoStop}%
\bibitem [{\citenamefont {Kiswandhi}\ and\ \citenamefont
  {Osada}(2021)}]{kiswandhi2021observation}%
  \BibitemOpen
  \bibfield  {author} {\bibinfo {author} {\bibfnamefont {A.}~\bibnamefont
  {Kiswandhi}}\ and\ \bibinfo {author} {\bibfnamefont {T.}~\bibnamefont
  {Osada}},\ }\href@noop {} {\bibfield  {journal} {\bibinfo  {journal} {Journal
  of Physics: Condensed Matter}\ }\textbf {\bibinfo {volume} {34}},\ \bibinfo
  {pages} {105602} (\bibinfo {year} {2021})}\BibitemShut {NoStop}%
\bibitem [{\citenamefont {He}\ \emph {et~al.}(2021)\citenamefont {He},
  \citenamefont {Isobe}, \citenamefont {Zhu}, \citenamefont {Hsu},
  \citenamefont {Fu},\ and\ \citenamefont {Yang}}]{he2021quantum}%
  \BibitemOpen
  \bibfield  {author} {\bibinfo {author} {\bibfnamefont {P.}~\bibnamefont
  {He}}, \bibinfo {author} {\bibfnamefont {H.}~\bibnamefont {Isobe}}, \bibinfo
  {author} {\bibfnamefont {D.}~\bibnamefont {Zhu}}, \bibinfo {author}
  {\bibfnamefont {C.-H.}\ \bibnamefont {Hsu}}, \bibinfo {author} {\bibfnamefont
  {L.}~\bibnamefont {Fu}}, \ and\ \bibinfo {author} {\bibfnamefont
  {H.}~\bibnamefont {Yang}},\ }\href@noop {} {\bibfield  {journal} {\bibinfo
  {journal} {Nature Communications}\ }\textbf {\bibinfo {volume} {12}},\
  \bibinfo {pages} {698} (\bibinfo {year} {2021})}\BibitemShut {NoStop}%
\bibitem [{\citenamefont {Kumar}\ \emph {et~al.}(2021)\citenamefont {Kumar}
  \emph {et~al.}}]{kumar2021room}%
  \BibitemOpen
  \bibfield  {author} {\bibinfo {author} {\bibfnamefont {D.}~\bibnamefont
  {Kumar}} \emph {et~al.},\ }\href@noop {} {\bibfield  {journal} {\bibinfo
  {journal} {Nat. Nanotechnol.}\ }\textbf {\bibinfo {volume} {16}},\ \bibinfo
  {pages} {421} (\bibinfo {year} {2021})}\BibitemShut {NoStop}%
\bibitem [{\citenamefont {Ho}\ \emph {et~al.}(2021)\citenamefont {Ho} \emph
  {et~al.}}]{ho2021hall}%
  \BibitemOpen
  \bibfield  {author} {\bibinfo {author} {\bibfnamefont {S.-C.}\ \bibnamefont
  {Ho}} \emph {et~al.},\ }\href@noop {} {\bibfield  {journal} {\bibinfo
  {journal} {Nat. Electron.}\ }\textbf {\bibinfo {volume} {4}},\ \bibinfo
  {pages} {116} (\bibinfo {year} {2021})}\BibitemShut {NoStop}%
\bibitem [{sm()}]{sm}%
  \BibitemOpen
  \href@noop {} {\bibinfo  {journal} {See the `Supplemental Materials' for more
  details on (i) the non-uniformity of relaxation time; (ii) spontaneous DC and
  AC Hall responses; (iii) the Hamiltonians used to produce the results.}\
  }\BibitemShut {NoStop}%
\bibitem [{Note1()}]{Note1}%
  \BibitemOpen
\bibfield  {journal} {  }\bibinfo {note} {In the limit where the relaxation
  time is a constant, $\tau (t) = \tau _*$, we observe that $J_y(t + \pi
  /\omega ) = J_y(t) + e^{-t/\tau _*} I(t)$. Thus, the contribution of $J_y(t)$
  to $\protect \bar J_y^{(2n+1)}$ is exponentially suppressed at $t \gg \tau _*
  \sim \tau _0$. Therefore, for a spatiotemporally uniform relaxation time only
  even harmonics of $J_y$ are appreciable deep in the steady state. It is
  easily checked that this non-perturbative outcome is consistent with the
  perturbative results obtained in the limit of a uniform $\tau $~\cite
  {nakai2019nonreciprocal,zhang2023}.}\BibitemShut {Stop}%
\bibitem [{\citenamefont {Morimoto}\ and\ \citenamefont
  {Nagaosa}(2018)}]{morimoto2018}%
  \BibitemOpen
  \bibfield  {author} {\bibinfo {author} {\bibfnamefont {T.}~\bibnamefont
  {Morimoto}}\ and\ \bibinfo {author} {\bibfnamefont {N.}~\bibnamefont
  {Nagaosa}},\ }\href@noop {} {\bibfield  {journal} {\bibinfo  {journal}
  {Scientific reports}\ }\textbf {\bibinfo {volume} {8}},\ \bibinfo {pages}
  {2973} (\bibinfo {year} {2018})}\BibitemShut {NoStop}%
\bibitem [{\citenamefont {Phong}\ and\ \citenamefont
  {Mele}(2023)}]{phong2023quantum}%
  \BibitemOpen
  \bibfield  {author} {\bibinfo {author} {\bibfnamefont {V.~T.}\ \bibnamefont
  {Phong}}\ and\ \bibinfo {author} {\bibfnamefont {E.}~\bibnamefont {Mele}},\
  }\href@noop {} {\bibfield  {journal} {\bibinfo  {journal} {Physical Review
  Letters}\ }\textbf {\bibinfo {volume} {130}},\ \bibinfo {pages} {266601}
  (\bibinfo {year} {2023})}\BibitemShut {NoStop}%
\bibitem [{\citenamefont {De~Beule}\ \emph {et~al.}(2024)\citenamefont
  {De~Beule}, \citenamefont {Gassner}, \citenamefont {Talkington},\ and\
  \citenamefont {Mele}}]{de2024floquet}%
  \BibitemOpen
  \bibfield  {author} {\bibinfo {author} {\bibfnamefont {C.}~\bibnamefont
  {De~Beule}}, \bibinfo {author} {\bibfnamefont {S.}~\bibnamefont {Gassner}},
  \bibinfo {author} {\bibfnamefont {S.}~\bibnamefont {Talkington}}, \ and\
  \bibinfo {author} {\bibfnamefont {E.}~\bibnamefont {Mele}},\ }\href@noop {}
  {\bibfield  {journal} {\bibinfo  {journal} {Physical Review B}\ }\textbf
  {\bibinfo {volume} {109}},\ \bibinfo {pages} {235421} (\bibinfo {year}
  {2024})}\BibitemShut {NoStop}%
\bibitem [{\citenamefont {Blinder}()}]{figure}%
  \BibitemOpen
  \bibfield  {author} {\bibinfo {author} {\bibfnamefont {S.~M.}\ \bibnamefont
  {Blinder}},\ }\href@noop {} {}\bibinfo {howpublished} {``Classical and
  Quantum Hall Effects'',
  \url{https://demonstrations.wolfram.com/ClassicalAndQuantumHallEffects},
  Wolfram Demonstrations Project}\BibitemShut {NoStop}%
\bibitem [{Note2()}]{Note2}%
  \BibitemOpen
  \bibinfo {note} {This limiting case also captures the essential aspects of
  higher dimensional metals because the components of momentum perpendicular to
  the applied field act like a label in the Boltzmann equation.}\BibitemShut
  {Stop}%
\bibitem [{Note3()}]{Note3}%
  \BibitemOpen
  \bibinfo {note} {Both $M_x$ and $M_y$ could be made $k_z$-dependent, but such
  a model would be unitarily equivalent to the one we write here.}\BibitemShut
  {Stop}%
\end{thebibliography}%

\clearpage
\appendix

\renewcommand{\thefigure}{S\arabic{figure}}
\setcounter{figure}{0}

\onecolumngrid

\section{Non-uniformity of relaxation time due to an  applied field} \label{sec:non-uniform}
In this section, through a self-consistent solution of the electron distribution function and  the relaxation time, we show that the latter develops a dependence on $\bs k$ and $\bs E$, even if it were a constant (i.e. $\tau_0$) in the unperturbed limit.
It is convenient to introduce 
\begin{align}
\gamma(\bs k, \bs E) = 1/\tau(\bs k, \bs E) 
\end{align}
to simplify the Boltzmann equation,
\begin{align}
& \qty(-\frac{e}{\hbar} \bs E) \cdot \grad_k f(\bs k, \bs E) + \gamma(\bs k, \bs E) [f(\bs k, \bs E) - f_0(\bs k)] = 0 \\
& \Rightarrow \hat E \cdot \grad_{\tilde k} g(\tilde{\bs k}, \bs E) - \frac{\gamma(\tilde{\bs k}, \bs E)}{\gamma_E} g(\tilde{\bs k}, \bs E) 
+ \hat E \cdot  \grad_{\tilde k} f_0(\tilde{\bs k})  = 0 
\end{align}
where $\tilde{\bs k} = \bs k / \Lambda $ with $\Lambda$ being the UV cutoff for $\bs k$, $g = f - f_0$, and 
\begin{align}
\gam_E = \frac{e E}{\hbar \Lambda}.
\end{align}
It would also be convenient to introduce the dimensionless quantities,
\begin{align}
\tilde \gamma(\tilde{\bs k}, \bs E) = \gamma(\tilde{\bs k}, \bs E)  \frac{\hbar}{\mu}, 
\qquad
\tilde \gamma_E = \gam_E  \frac{\hbar}{\mu} = \frac{eE}{\mu \Lam}.
\end{align}
Assuming $\bs E = E \hat x$,  the Boltzmann equation can be expressed in terms of dimensionless quantities,  
\begin{align}
\partial_{\tilde k_x} g(\tilde{\bs k}, E) - \frac{\gamma(\tilde{\bs k}, E)}{\gamma_E} g(\tilde{\bs k}, E) 
+ \partial_{\tilde k_x}f_0(\tilde{\bs k})  = 0. 
\label{eq:BE}
\end{align}
Being a linear differential equation, \eqref{eq:BE} can be solved exactly with the boundary condition $g(\tilde k_x = 1, \tilde k_y, E) = 0$ (assuming $E>0$) to obtain
\begin{align}
g(\tilde{k}_x, \tilde k_y, E) = \exp{- \frac{1}{\tilde \gam_E} \int_{\tilde k_x}^1 \dd{w} \tilde \gamma(w, \tilde k_y,  E) } \int_{\tilde k_x}^1 \dd{z} \exp{\frac{1}{\tilde \gam_E} \int_{z}^1 \dd{w} \tilde \gamma(w, \tilde k_y,  E) } ~\partial_{z}f_0(z, \tilde k_y).
\label{eq:full_g}
\end{align}

From the collision integral we obtain
\begin{align}
\gam(\bs k, \bs E) g(\bs k, \bs E) &= \frac{2\pi}{\hbar} \int \dd{\bs k'} \delta(\veps_k - \veps_{k'}) ~|U(\bs k- \bs k')|^2 \qty[g(\bs k, \bs E) - g(\bs k', \bs E)] \\ 
\Rightarrow \tilde \gam(\tilde{\bs k}, \bs E)&= \frac{2\pi \Lambda^2}{\mu} \int \dd{\tilde{\bs k}'} \delta(\veps_{\tilde k} - \veps_{\tilde k'}) ~|U(\tilde{\bs k}- \tilde{\bs k}')|^2 \qty[1 - \frac{g(\tilde{\bs k}', \bs E)}{g(\tilde{\bs k}, \bs E)}] \label{eq:full_gam}
\end{align}
In principle one can solve the coupled equations \eqref{eq:full_g} and \eqref{eq:full_gam} to determine the two unknowns, $\gam(\bs k, \bs E)$ and $g(\bs k, \bs E)$

\paragraph*{~~~} ~~~ \\

In order to understand the structure of the solution, let us consider a 1D metal with a parabolic band~\footnote{This limiting case also captures the essential aspects of higher dimensional metals because the components of momentum perpendicular to the applied field act like a label in the Boltzmann equation.}.
In this case,
\begin{align}
\veps_k = \frac{\hbar^2 k_x^2}{2 m} = \veps_\Lambda \tilde k_x^2,
\end{align}
which simplifies \eqref{eq:full_gam} to 
\begin{align}
\gam(\tilde k_x, E) &= \frac{\Lambda}{\hbar \veps_\Lambda} \int_{-1}^1 \dd{\tilde k_x'} \delta(\tilde k_x'^2  -  \tilde k_x^2) ~|U(\tilde k_x - \tilde k_x')|^2 \qty[1 - \frac{g(\tilde k_x', E)}{g(\tilde k_x, E)}] \\
&= \frac{\Lambda}{2\hbar \veps_\Lambda |\tilde k_x|} \qty[
|U(0)|^2 \times 0
+ |U(2 \tilde k_x)|^2 \qty{1 - \frac{g(-\tilde k_x, E)}{g(\tilde k_x, E)} }
] 
\end{align}
which implies
\begin{align}
\tilde \gam(\tilde k_x, E) = \frac{\Lambda |U(2 \tilde k_x)|^2}{2\mu \veps_\Lambda |\tilde k_x|} 
 \qty[1 - \frac{g(-\tilde k_x, E)}{g(\tilde k_x, E)} ].
\label{eq:gamma1D}
\end{align}
Since the shift of the Fermi sea  due to the applied field breaks the $k_x \mapsto - k_x$ symmetry, $g(-\tilde k_x, E) \neq  \pm g(\tilde k_x, E)$, which implies that $\gam(\tilde k_x, E)$ does not possess a fixed parity under space-inversion.

The change in the distribution function takes the form,
\begin{align}
g(\tilde{k}_x, E) = \exp{- \frac{1}{\tilde \gam_E} \int_{\tilde k_x}^1 \dd{w} \tilde \gamma(w,  E) } \int_{\tilde k_x}^1 \dd{z} \exp{\frac{1}{\tilde \gam_E} \int_{z}^1 \dd{w} \tilde \gamma(w,   E) } ~\partial_{z}f_0(z).
\label{eq:g1D}
\end{align}
Because $g$ does not have a fixed parity under $k_x \mapsto - k_x$, it is difficult to analytically  obtain a self-consistent solution for $\gamma$ with the most  general possible momentum-dependence.
Nonetheless, self-consistent solutions can be obtained numerically by employing ansatz forms of $\gamma$.
In the remainder of this section we will discuss these solutions both at $T=0$ and for $T > 0$.

\subsection*{Ansatz \# 1}
We use the simplest possible non-trivial form for $\gamma$,
\begin{align}
\gamma(\tilde k_x, E) = \Theta(\tilde k_x) \gam_+(E) + \Theta(-\tilde k_x) \gam_-(E) \equiv \gamma_0(E) + \mbox{sign}(k_x) \gamma_1(E).
\end{align}
Here, the two equivalent expressions are related as  $\gam_\pm(E) = \gam_0(E) \pm \gam_1(E) \Rightarrow \gam_{0,1}(E) = \frac{1}{2} \qty[\gam_+(E) \pm \gam_-(E)]$.
In standard linear-response calculations, because one seeks to understand the change with respect to a state at global equilibrium, one sets $\gam_+(E) = \gam_-(E) = 1/\tau_0$, where $1/\tau_0$
 is the relaxation time in the absence of applied field.
Here, we shall solve for $\gam_+(E)$ and $\gam_-(E)$ by evaluating \eqref{eq:full_gam} at $\tilde k_x = \pm \tilde k_F$ (this is sufficiently general  because the ansatz is only sensitive to the sign of $k_x$).
Therefore,
\begin{align}
& \tilde \gam_+(E) = \tilde \gam(\tilde k_F, E) = \frac{\Lambda |U(2 \tilde k_F)|^2}{2\mu \veps_\Lambda \tilde k_F} 
 \qty[1 - \frac{g(-\tilde k_F, E)}{g(\tilde k_F, E)} ] \nn \\
& \tilde \gam_-(E) = \tilde \gam(-\tilde k_F, E) = \frac{\Lambda |U(-2 \tilde k_F)|^2}{2\mu \veps_\Lambda \tilde k_F} 
 \qty[1 - \frac{g(\tilde k_F, E)}{g(-\tilde k_F, E)} ],
\label{eq:gamPM}
\end{align}
which leads to 2 unknowns ($\tilde \gam_\pm$) to be determined from 2 equations, at a fixed $E$.

\begin{figure}[!t]
\centering
\subfloat[]{%
\includegraphics[width=0.475\columnwidth]{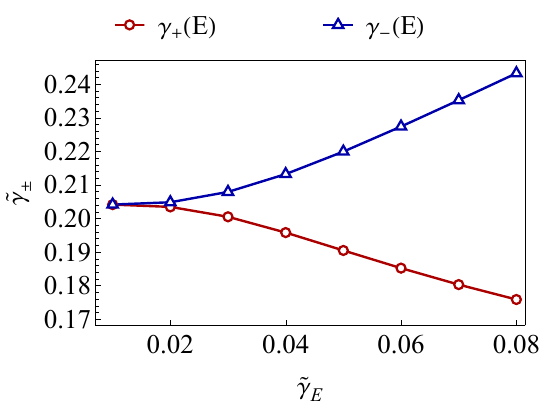}
}
\hfill
\subfloat[]{%
\includegraphics[width=0.475\columnwidth]{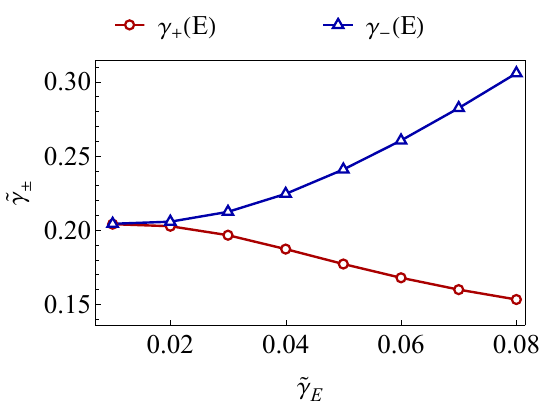}
}
\caption{$\tilde \gam_\pm$ as a function of $E$. 
(a) At $T=0$ we obtain $\tilde \gam_\pm(E)$ by  numerically solving the coupled pair of equations in \eqref{eq:gamPMT0}. 
(b) At $T > 0$ $\gam_\pm(E)$ is obtained by a self-consistent solution to \eqref{eq:g1D} and \eqref{eq:gamPM}.
Here, $\tilde k_F = \sqrt{0.06}$,  $k_B T  = 10^{-3} \veps_\Lam$, and $\frac{\Lambda |U(\pm 2 \tilde k_F)|^2}{2\mu  \veps_\Lambda \tilde k_F} = 0.025$.
We note that $\gam_\pm$ are both $T$ and $\mu$ dependent, and should be expressed as $\gam_\pm(k_x, E, T, \mu)$.
}
\label{fig:gamPM}
\end{figure}

At $T = 0$, it is possible to obtain an analytic expression for $g$,
\begin{align}
g(\tilde k_x, E) = \begin{cases}
    -\exp{\frac{\tilde \gam_+}{\tilde \gam_E}(\tilde k_x - \tilde k_F)},  & \tilde k_F \geq \tilde k_x \geq 0 \\  
    -\exp{\frac{1}{\tilde \gam_E}(\tilde \gam_- \tilde k_x - \tilde \gam_+ \tilde k_F)},  & 0 > \tilde k_x > - \tilde k_F \\ 
    \exp{\frac{\tilde \gam_-}{\tilde \gam_E}(\tilde k_x + \tilde k_F) } - \exp{\frac{1}{\tilde \gam_E}(\tilde \gam_- \tilde k_x - \tilde \gam_+ \tilde k_F) },  & - \tilde k_F \geq \tilde k_x \geq -1 \\
    0, & \mbox{otherwise}
\end{cases}
\end{align}
where we have utilized the fact that $\lim_{T \to 0} f_0(k_x) = \Theta(\mu - \veps_k) \equiv \Theta(\tilde k_F - |\tilde k_x|)$ with $\tilde k_F = \sqrt{\mu/\veps_\Lam}$.
It is straightforward to determine that 
\begin{align}
& g(\tilde k_F, E) = -1 \\
& g(-\tilde k_F, E) = 1- \exp{-\frac{\gam_+ + \gam_-}{\gam_E} \tilde k_F},
\end{align}
We note that this matches the $T \to 0$ limit of the solution of $g$ at finite $T$.
Thus, we obtain 
\begin{align}
& \tilde \gam_+(E) =  \frac{\Lambda |U(2 \tilde k_F)|^2}{2\mu  \veps_\Lambda \tilde k_F} 
~\qty[2 - \exp{-\frac{\gam_+ + \gam_-}{\gam_E} \tilde k_F}], \nn \\
& \tilde \gam_-(E) = \frac{\Lambda |U(-2 \tilde k_F)|^2}{2\mu \veps_\Lambda \tilde k_F} 
~\frac{2 - \exp{-\frac{\gam_+ + \gam_-}{\gam_E} \tilde k_F}}{1 - \exp{-\frac{\gam_+ + \gam_-}{\gam_E} \tilde k_F}}.
\label{eq:gamPMT0}
\end{align}
This pair of equations can now be solved numerically to obtain $\gam_{\pm}(E)$ [see Fig.~\ref{fig:gamPM}], which in turn determines the self-consistent solution for $g(\tilde k_x, E)$.

At $T>0$ the computation is carried out numerically by noting that
\begin{align}
& f_0(k_x) = \qty[\exp{\beta (\veps_k - \mu)} + 1]^{-1} 
= \qty[\exp{\underbrace{\beta \veps_\Lam}_{\displaystyle{\tilde \beta}} (\tilde k_x^2 - \tilde k_F^2)} + 1]^{-1} \equiv f_0(\tilde k_x) \\ 
\Rightarrow & \partial_{\tilde k_x} f_0(\tilde k_x) = - 2\tilde \beta \tilde k_x ~\frac{\exp{\tilde \beta  (\tilde k_x^2 - \tilde k_F^2)}}{\qty[\exp{\tilde \beta (\tilde k_x^2 - \tilde k_F^2)} + 1]^2}.
\end{align}
The result is presented in Fig.~\ref{fig:gamPM}.
Numerically fitting the results implies $\gam_\pm \equiv \gam_{0,1}$ have the form
\begin{align}
\gam_{0}(E) = \sum_{j=0}^4 a_j E^j \qquad 
\gam_{1}(E) = \sum_{j=1}^4 b_j E^j.
\end{align}
We note that the parameters $a_j$ and $b_j$ in general depends on $T$, $\mu$ and other model parameters.
Therefore, we conclude, generally  
\begin{align}
\gam(\tilde k_x, E) \neq \gam(-\tilde k_x, E).
\end{align}

\subsection*{Ansatz \# 2}
\begin{figure}[!t]
\centering
\subfloat[]{%
\includegraphics[width=0.475\columnwidth]{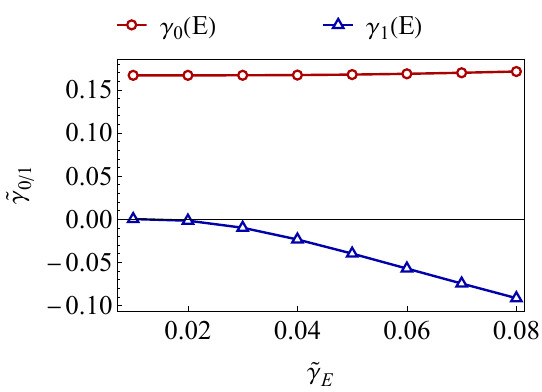}
}
\hfill
\subfloat[]{%
\includegraphics[width=0.475\columnwidth]{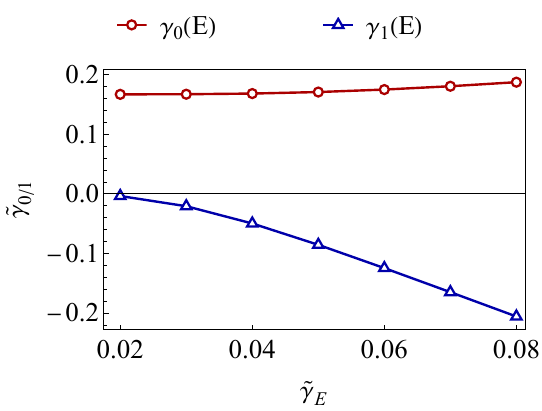}
}
\caption{$\gam_{0/1}$ as a function of $E$. 
At (a) $T=0$; (b) $k_B T = 10^3 \veps_\Lam$. 
Here, $\tilde k_F = 0.3$,   and $\frac{\Lambda |U(\pm 2 \tilde k_F)|^2}{2\mu  \veps_\Lambda \tilde k_F} = 0.025$.
We note that $\gam_{0/1}$ are both $T$ and $\mu$ dependent, and should be expressed as $\gam_{0/1}(k_x, E, T, \mu)$.
}
\label{fig:gam01}
\end{figure}

In order to ensure the above behavior was not special for the chosen ansatz, here, we repeat the calculation with an  alternative ansatz, 
\begin{align}
\gam(\tilde k_x, E) = \gam_0(E) + \gam_1(E) \tilde k_x.
\end{align}
We follow the same steps as the previous ansatz and obtain the results presented in Fig.~\ref{fig:gam01}.
Here, too, we conclude that $\gam(\tilde k_x, E) \neq \gam(-\tilde k_x, E)$ for $T \geq 0$.

\section{Further details on  spontaneous DC Hall response}
In this section we present further details on the breakdown of the perturbative expansion in the electric field and obtain the expression of the change of the electron distribution function in the $T \to 0$ limit.

\subsection{Breakdown of perturbative expansion in the DC limit}
In this section we demonstrate the breakdown of the perturbative expansion in $k_E$ of the anomalous Hall current in the DC limit.
For the leading order non-vanishing term  in the expansion (say, at order $n = n_0$) to be a good approximation to $J_y(\bs E)$, the electric field has to be sufficiently weak such that 
\begin{align}
\tilde k_E^{n} \qty|\int_{\bs k}  f_0~ \partial_{ \tilde k_x}^{n} \Om_{xy}| \ll \tilde k_E^{n_0} \qty|\int_{\bs k}  f_0~ \partial_{ \tilde k_x}^{n_0} \Om_{xy}|
\end{align}
for all non-vanishing terms at order $n > n_0$.
This implies
\begin{align}
\tilde k_E^{n - n_0} \ll \frac{\qty|\int_{\bs k}  f_0~ \partial_{ \tilde k_x}^{n_0} \Om_{xy}|}{\qty|\int_{\bs k}  f_0~ \partial_{ \tilde k_x}^{n} \Om_{xy}|} \sim \tilde k_\bc^{n-n_0}.
\end{align}
Therefore, only in the limit $\eps   \ll 1$, $J_y(\bs E)$ satisfies the leading-order scaling obtained by the perturbative expansion in $\tilde k_E$.

\begin{figure}[!t]
\centering
\includegraphics[width=0.5\columnwidth]{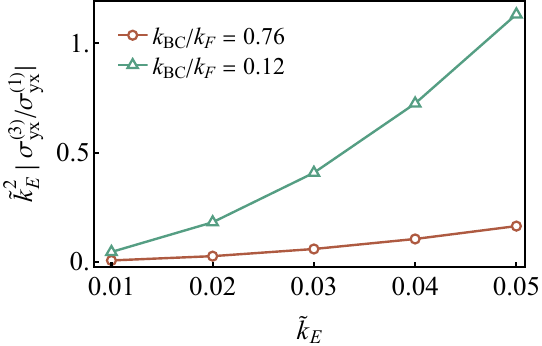}
\caption{Breakdown of the perturbative expansion in $E = |\bs E|$ with increasing proximity of the extrema of the Berry curvature density, $\Om_{xy}$, to the Fermi surface. 
As $k_\bc/k_F$ is reduced, the magnitude of the formally higher order terms in Eq.~(5) become comparable to the leading order term.
Here, $\sig_{yx}^{(n)}$ is the $n$-th order term on the right-hand side of Eq.~(5) of the main text expressed in units of $E$.
We have used the parameter $w$ in Eq.~\eqref{eq:dimless}  and the chemical potential, $\mu$, to tune $k_\bc$ at a fixed $k_F$ (`$k_F$' is defined as half the diameter of individual Fermi pockets at $k_y = 0$; c.f. Fig.~1 of the main text.).
}
\label{fig:breakdown}
\end{figure}

With the help of the model in the main text, in Fig.~\ref{fig:breakdown} we show that the breakdown happens at progressively weaker fields as the extrema of the Berry curvature draws nearer to the Fermi surface. 
In particular, the ratio between the  sub-leading and leading terms in the expansion in Eq.~(5) of main text grows faster with the applied field ($k_E$), as $k_\bc$ is reduced at a fixed $k_F = 0.1 \Lam$.

\subsection{Zero-temperature limit of DC response}
We begin by noting that in the finite-$T$ expression of $g$ in the main text the $\partial_{\tilde q} f_0$ term in the integrand is a result of the fact that we expressed $f = f_0 + g$. 
Its presence only represents this mathematically convenient decomposition of $f$, since away from the weak-field regime, which we consider here, $g$ is not small compared to $f_0$. 
Therefore, $g$ receives contributions from states away from the equilibrium Fermi surface, which, in fact, play an important role in the non-perturbative response discussed here. 
We better appreciate this point, we send $T \to 0$.

In the $T \to 0$ limit 
$f_0(\bs k) \to \Theta(\mu - \veps_k)$ with $\mu$, $\veps_k$, and $\Theta(x)$ being the chemical potential, dispersion of the partially 
filled band, and the Heaviside theta function, respectively.
The general solution of $g$ in the $T\to 0$ can be expressed in a closed form,
\begin{align}
\lim_{T\to 0} g(\bs k, \bs E) =  - \sum_{n=1}^{N_y} & \Theta(k_{F,x}^{(n)} - k_x)~ e^{-(k_{F,x}^{(n)} - k_x)/k_E} ~\mbox{sgn}\qty{v_x(\bs k_F^{(n)}) }.
\label{eq:g-T=0}
\end{align}
Here, $N_y$ is the number of times the $k_x$-axis intersects with  the Fermi surface for a fixed $k_y$, $\bs k_F^{(n)}$'s locate these intersections, and $v_x = \partial_{k_x} \veps_k$.
The exponential factor in \eq{eq:g-T=0} limits the change in the occupation number to a region of width $\sim k_E$ about $\bs k_F^{(n)}$.
This establishes two regimes that would be present in any response function: for $k_E \ll |\bs k_F|$ ($k_E \gtrsim |\bs k_F|$) primarily states on the Fermi surface (states both at the Fermi surface and in the Fermi sea) contribute to transport.
The sign of the $x$-component of the group velocity dictates the manner in which the Fermi sea shifts, with a positive (negative) sign implying a depletion (replenishment) of states.
Clearly, as $k_E$ increases, the peak in $g$ becomes broad. 
This broadening increases further as $T$ increases. 

\section{Aspects of  spontaneous AC Hall  response}
In this section we (i) obtain the perturbative expressions of the harmonics of $g$ for a general $\bs k$ and $\bs E$ dependent $\tau$; (ii) demonstrate the necessity of the method of solution of $f$ used in the main text. 
We  write the time-dependent Boltzmann equation as,
\begin{align}
& \partial_t f(t, \bs k, \bs E) + \qty[-\frac{e}{\hbar} \bs E(t)] \cdot \grad_k f(t, \bs k, \bs E) + \gamma(t, \bs k, \bs E) [f(t, \bs k, \bs E) - f_0(\bs k)] = 0 \\
& \Rightarrow \frac{1}{\gam_E} \partial_{t} g(t, \tilde{\bs k},  E) -  \cos{\om t} ~\hat E \cdot \grad_{\tilde k} g(t, \tilde{\bs k}, E) + \frac{\gamma(t, \tilde{\bs k},  E)}{\gamma_E} g(t, \tilde{\bs k}, E) 
= \cos{\om t} ~\hat E \cdot  \grad_{\tilde k} f_0(\tilde{\bs k})  
\end{align}
where $\tilde{\bs k} =  \bs k / \Lambda $ with $\Lambda$ being the UV cutoff for $\bs k$
Henceforth, we shall set $\hat E = \hat x$.

Introducing the Fourier components through 
\begin{align}
& g(t, \tilde{\bs k}, E) = \sum_{n = 0} \bar g_n(\tilde{\bs k}, E) ~ \cos{ n \om t} \\ 
& \gam(t, \tilde{\bs k}, E) = \sum_{n = 0} \bar \gam_n(\tilde{\bs k}, E) ~ \cos{n \om t}
\label{eq:gamma-fourier}
\end{align}
we obtain
\begin{align}
& - \frac{1}{\gamma_E} \sum_{n = 0}  (n \om) \sin{(n \om t)} \bar g_n(\tilde{\bs k}, E) 
+ \frac{1}{\gamma_E} \sum_{n, m =0} \cos{(n \om t)} \cos{(m \om t)} \bar \gam_n(\tilde{\bs k}, E) \bar g_m(\tilde{\bs k}, E)  \nn \\
& \qquad = \cos{\om t} ~\partial_{\tilde k_x} f_0(\tilde{\bs k}) 
+ \sum_{n = 0} \cos{(\om t)} \cos{(n \om t)} ~\partial_{\tilde k_x} \bar g_n(\tilde{\bs k}, E).
\end{align}
Matching the coefficients of $e^{i n \om t}$ we obtain the following equations (we suppress the dependence on $\tilde{\bs{k}}, E$ for notational convenience)
\begin{align}
e^{0}: &\quad 
\bar \gam_0 \bar g_0
+  \frac{1}{2} \sum_{n=1} \bar \gam_n \bar g_n = \frac{\gam_E}{2} \partial_{\tilde k_x} \bar g_1 \nn \\
e^{i\om t}: &\quad 
i \om \bar g_1 +  \qty[
\bar \gam_1 \bar g_0 + \bar \gam_0 \bar g_1 + \half \sum_{n=1} \qty(\bar \gam_{n+1} \bar g_n + \bar \gam_n \bar g_{n+1} )
] = \frac{\gam_E}{2} \qty[
2\partial_{\tilde k_x} f_0 
+ 2\partial_{\tilde k_x} \bar g_0 
+ \partial_{\tilde k_x} \bar g_2 
] \nn \\ 
e^{2 i\om t}: &\quad 
2  i \om \bar g_2 +  \qty[
\bar \gam_2 \bar g_0 + \bar \gam_0 \bar g_2 + \half \bar \gam_1 \bar g_1 +  \half \sum_{n=1} \qty(\bar \gam_{n+2} \bar g_n + \bar \gam_n \bar g_{n+2} )
] = \frac{\gam_E}{2} \qty[
\partial_{\tilde k_x} \bar g_1 
+ \partial_{\tilde k_x} \bar g_3 
] \nn \\ 
\vdots \quad  & ~ \nn \\
e^{i m \om t}: &\quad 
 i m \om \bar g_m +  \qty[
\bar \gam_m \bar g_0 + \bar \gam_0 \bar g_m + \half \sum_{n=1}^{m-1} \bar \gam_{n} \bar g_{m-n} + \half \sum_{n=1} \qty(\bar \gam_{n+m} \bar g_n + \bar \gam_n \bar g_{n+m} )
] \nn \\ 
& \quad ~ = \frac{\gam_E}{2} \qty[
\partial_{\tilde k_x} \bar g_{m-1} 
+ \partial_{\tilde k_x} \bar g_{m+1} 
].
\label{eq:coupled-g}
\end{align}

\subsection{Uniform relaxation time approximation}
The conventional perturbative results for up to $2\omega$ AC Hall response~\cite{sodemann2015}  can be obtained by setting all $\bar g_{n\geq 3} = 0$, and $\bar \gamma_{n \geq 1} = 0$ with the assumption that these produce higher-order $E$ dependence.
Further, we need to set $\bar \gam_0 = 1/\tau_0$.
Thus from the coefficients of $e^0$,  
$e^{2 i\om t}$, and $e^{i\om t}$, respectively, we obtain
\begin{align}
& \bar g_0 = \frac{(e E/\hbar)}{2/\tau_0} 
\partial_{k_x} \bar g_1 \nn \\
& \bar g_2  = \frac{(e E/\hbar)}{2(2i\omega + 1/\tau_0)} 
\partial_{k_x} \bar g_1 \nn \\
& \bar g_1 =  \frac{(e E/\hbar)}{2(i\omega + 1/\tau_0) } \qty[
2\partial_{k_x} f_0 
+ 2\partial_{k_x} \bar g_0 
+ \partial_{k_x} \bar g_2 
].
\label{eq:gs-1}
\end{align}
Since $\bar g_{n\neq 1} \propto E \partial_{k_x} \bar g_1$, we can drop the last two terms on the right hand side of the expression of $\bar g_{n\neq 1}$ in the interest of obtaining the leading order terms in $E$, which leads to the well-known solution,
\begin{align}
& \bar g_0 = \frac{k_E^2}{2(i\omega \tau_0 + 1)} \partial_{k_x}^2 f_0 \nn \\
& \bar g_1 = \frac{k_E}{(i\omega \tau_0 + 1) } 
\partial_{k_x} f_0 \nn \\ 
& \bar g_2 = \frac{k_E^2}{2(i\omega \tau_0 + 1)(2i\omega \tau_0 + 1)} \partial_{k_x}^2 f_0,
\end{align}
where $k_E = e E \tau_0/\hbar$ was introduced in the main text.
Since both $\bar g_0$ and $\bar g_2$ are even function of $\bs k$, they do not contribute to the Hall current.
This leads to the vanishing of the $1\om$ response. 

\subsection{Non-uniform relaxation time}
Motivated by the results in Section~\ref{sec:non-uniform}, we employ a relaxation time whose inverse is of the form
\begin{align}
\frac{1}{\tau(\bs k, \bs E)} = \gamma(\bs k, \bs E) = \sum_{j=0}^N a_j E^j + \mbox{sign}(\hat E \cdot \bs k) \sum_{j=1}^N b_j E^j,
\end{align}
where $\bs E = E \hat E$ and $a_0 = 1/\tau_0$. 
To obtain the expression of $\gamma$ in the AC limit we generalize $E \to E \cos{(\om t)}$. 
This results in the expression in Eq.~\eqref{eq:gamma-fourier}.
For extracting the leading order behaviors henceforth we set $N=2$.
Up to this order we find 
\begin{align}
\bar \gam_0 = a_0 + \frac{1}{2}\qty{a_2 + b_2 \mbox{sign}(\hat E \cdot \bs k)} E^2; \qquad
\bar \gam_1 = \qty[ a_1 + b_1 \mbox{sign}(\hat E \cdot \bs k) ] E; \qquad
\bar \gam_2 = \frac{1}{2}\qty[ a_1 + b_1 \mbox{sign}(\hat E \cdot \bs k) ] E^2.
\end{align}
We note that, in general, $\bar \gam_j$'s are polynomials in $E$ and contains $\order{E^3}$ terms.
However, these higher order terms are not important for our present objective.

Next, we solve for $\bar g_{j\leq 2}$ from Eq.~\ref{eq:coupled-g} with  $\bar \gam_{j\leq 2}$ finite to obtain the generalization to the expression in Eq.~\ref{eq:gs-1} (note that, here, we first set $N = 4$, then obtain the solution, and, finally, truncate the solution to the order indicated below), 
\begin{align}
& \bar g_0 = \frac{(e E/\hbar)}{2/\tau_0} \partial_{k_x} \bar g_1
-  \frac{(e E/\hbar) \bar \gam_1}{(2/\tau_0)(i\om + 1/\tau_0)}
\qty[
\partial_{k_x} f_0 
+ \partial_{k_x} \bar g_0 
+ \frac{1}{2} \partial_{k_x} \bar g_2 ] \nn \\
& \qquad \qquad + \frac{\qty[(4i\om + 3/\tau_0)\bar \gam_1^2 - 2(i\om + 1/\tau_0) \bar \gam_2/\tau_0 ] (e E/\hbar) }{2(2/\tau_0)^2 (i\om + 1/\tau_0)(2i\om + 1/\tau_0)} \partial_{k_x} \bar g_1 
+ \order{E^5} \nn \\
& \bar g_1 = \frac{(e E/\hbar)}{(i\omega + 1/\tau_0) } \qty[
\partial_{k_x} f_0 
+ \partial_{k_x} \bar g_0 
+ \frac{1}{2} \partial_{k_x} \bar g_2 ] 
- \frac{ (3/\tau_0 + 4 i \om) (e E/\hbar) \bar{\gam}_1}{(4/\tau_0)(i\om + 1/\tau_0)(2i\om + 1/\tau_0)} \partial_{k_x} \bar g_1 \nn \\
& \qquad \qquad + \frac{\qty[(4i\om + 3/\tau_0)\bar \gam_1^2 - 2(2i\om + 1/\tau_0)\bar \gam_2/\tau_0] (eE/\hbar)}{(4/\tau_0)(i\om + 1/\tau_0)^2 (2i\om + 1/\tau_0)} \partial_{k_x}  f_0 
+ \order{E^5}
\nn \\
& \bar g_2 = \frac{(e E/\hbar)}{2(2i\omega + 1/\tau_0)} 
\partial_{k_x} \bar g_1 
- \frac{(e E/\hbar) \bar \gam_1}{2(i\om + 1/\tau_0)(2i\om + 1/\tau_0)(3i\om + 1/\tau_0)} \qty[(3i\om + 1/\tau_0) \qty(\partial_{k_x}  f_0 + \partial_{k_x} \bar g_0) + (2i\om + 1/\tau_0) \partial_{k_x} \bar g_2] \nn \\
& \qquad \qquad + \frac{\qty[(3i\om 2/\tau_0) \bar \gam_1^2 - 2(i\om + 1/\tau_0)(3i\om + 1/\tau_0)\bar \gam_2] (e E/\hbar)}{(4/\tau_0)(i\om + 1/\tau_0)(2i\om + 1/\tau_0)(3i\om + 1/\tau_0)} \partial_{k_x} \bar g_1 + \order{E^5}.
\end{align}
We observe that, given the $\bs k$-dependence of $\bar \gamma_j$'s, the $\bar g_j$'s no longer have a fixed parity under space-inversion.
Importantly, both $\bar g_0$ and $\bar g_2$ now have terms that are odd in $\bs k$, which in turn will lead to a finite $1\omega$ response.
For the form of non-uniform $\tau$ considered in the main text, all $a_{j\geq 1} = 0$ but $b_j$'s are finite. 
Therefore, only the third terms in the expressions of $\bar g_0$ and $\bar g_2$ contribute to the Hall current through the $\bar \gam_2$ dependence.
Since this term $\sim E^4$, the $1\om$ component of the Hall current will scale as $E^5$, as obtained through the non-perturbative solution presented in the main text. 

\subsection{Non-perturbative solution in $E$}
The above process fails beyond the asymptotically weak-$E$ limit, even if we set $\gam(t, \tilde{\bs k}, \bs E) = 1/\tau_0$, due to the coupling among all $n\omega$ harmonics of $g$ [this can be readily deduced by attempting to solve for $\bar g_n$'s via  Eq.~\ref{eq:coupled-g}].
Thus, a general solution for $\bar g_n$ is inaccessible if we Fourier transform $g$ before solving the Boltzmann equation.
This motivates the strategy for the non-perturbative solution that is discussed in the main text, where we first solve for $f$ non-perturbatively in $E$ and then Fourier transform the transverse current to determine the $n\omega$ response.


\section{More about the model in Eq. (8)}
We recall the two-band $k.p$ model of an two dimensional inversion-broken  insulator,
\begin{align}
\mf{h}(\bs k) =&   \qty(M_x - B_x k_x^2 ) \sig_1 + A k_y \sig_2 + \qty(M_y - B_y k_y^2 ) \sig_3\nn \\
& + B_0 k_x^2 \sigma_0,
\end{align}
where $\qty{A, M_j, B_j}$ are momentum-independent parameters for the model.
Since $\mf h^*(- \bs k) = \mf{h}(\bs k)$ and $\mf{h}(-k_x, k_y) = \mf{h}(k_x, k_y)$, the Hamiltonian preserves TRS with $\mc T^2 = +1$ and the $x$-mirror symmetry.
It, however, breaks 2D inversion or $\mathcal C_2$ symmetry.
Thus, upon doping, the 2D insulator described by $\mf{h}(\bs k)$ would be expected to support a non-linear Hall response~\cite{sodemann2015}.
The parameter $M_y$ tunes a topological phase transition between two insulating states with the critical point located at $M_y = 0$.
At the critical point, a pair of linear band-crossings appears on the $k_x$-axis, where the Berry curvature  diverges as $\Om_{xy} \sim |\delta \bs k|^{-1}$ with $\dl \bs k$ being  the deviation from a band-crossing point.

In order to obtain a well-defined effective model, we introduce an ultraviolet (UV) momentum cutoff, $\Lambda$, which allows us to write $k_j = \Lambda \tilde k_j$ with $\tilde k_j$ being dimensionless, and define  other dimensionless parameters,
\begin{align}
\qty{w, b, a, \delta, \Delta} = \qty{\frac{B_0}{B_x}, \frac{B_y}{B_x}, \frac{A}{B_x \Lambda}, \frac{M_x}{B_x \Lambda^2}, \frac{M_y}{B_x \Lambda^2}},
\label{eq:dimless}
\end{align}
such that the Hamiltonian is expressed in terms of dimensionless quantities,
\begin{align}
\mf{h}(\bs k)  =&   E_\Lambda \qty[\qty(\delta -  \tilde k_x^2 ) \sig_1 + a \tilde k_y \sig_2 + \qty(\Delta - b \tilde{k}_y^2 ) \sig_3 
+ w \tilde{k}_x^2 \sigma_0],
\end{align}
with $E_\Lambda \coloneqq B_x \Lambda^2$ being the UV energy cutoff and sets the overall energy scale for the model.
The gap between the two bands is controlled by $|1 - \delta/\Delta|$, which may be considered as a ``correlation length'' that controls the transition between two topological insulating states lying on either side of $\delta/\Delta = 1$.
Here, we will assume the ratio $\delta/\Delta > 0$ to be sufficiently smaller than $1$, such that the band-gap is the largest scale, and all inter-band effects are suppressed.  

\begin{figure}[!]
\centering
\includegraphics[width=0.5\linewidth]{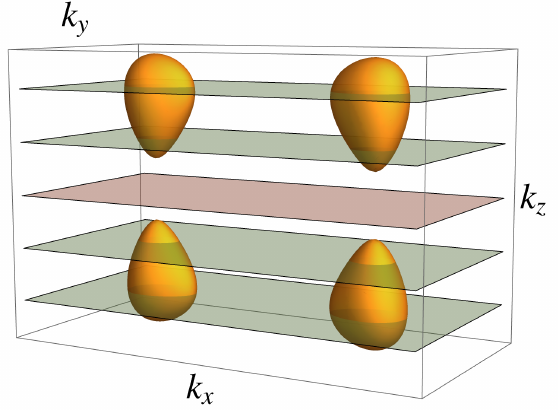}
\caption{Connection between three dimensional, non-centrosymmetric,  doped Weyl semimetals and two dimensional metals. 
The (orange) spheroid pockets are Fermi surfaces enclosing the four Weyl points. 
The $k_z$ planes marked in green are metallic, and their representative contribution to non-linear Hall response was studied in this paper. 
The $k_z$ plane in red represents two dimensional insulators which do not contribute to non-linear Hall response.}
\label{fig:slices}
\end{figure}

\section{Anomalous Hall response in $\mc C_4 \mc T$ symmetric metal}
We consider the metallic state obtained by electron-doping the 2D insulator described by
\begin{align}
H = t_1 \cos\frac{k_x}{2}\cos\frac{k_y}{2} \sigma_1 + t_2 \sin\frac{k_x}{2}\sin\frac{k_y}{2} \sigma_2 + t_3(\cos{k_x}+\cos{k_y} - \Delta)\sigma_3
+ t_0 (\cos{k_x}+\cos{k_y}) \sigma_0.
\end{align}
This Hamiltonian is invariant under $\mc C_4 \mc T$ transformation  under which $H(k_x, k_y) \mapsto H^*(k_y, -k_x) = H(k_x, k_y)$.
This leads to the vanishing of both the Chern number for each band, and the dipole contribution to the Hall current once doped.

We Taylor expand $H$ about the origin up to order $|\delta \bs k|^2$, and  use the following parameters
\begin{align}
\Delta = 1.98; \quad t_1/t_3 = 0.05; \quad t_2/t_3 = 0.5; \quad t_0/t_3 = 0.1; \quad \mu = 0.055
\end{align}
to obtain the result in the main text.

\section{Connection between non-centrosymmetric Weyl semimetal and the 2D model}
It is straightforward to generalize the model in Eq.~(8) of the main text to describe a TRS-preserving non-centrosymmetric Weyl semimetal (WSM) by $M_y \to M_z - B_z k_z^2$~\footnote{Both $M_x$ and $M_y$ could be made $k_z$-dependent, but such a model would be unitarily equivalent to the one we write here.}.
Thus, the Hamiltonian of the WSM is
\begin{align}
H_{\text{WSM}} = (M_x - B_x k_x^2) \sigma_1 + A k_y \sigma_2 + (M_z - B_z k_z^2 - B_y k_y^2) \sigma_3 + B_0 k_x^2 \sigma_0.
\end{align}
It supports 4 Weyl points on the $k_y = 0$ plane at $(k_x, k_z) = \qty(\pm \sqrt{M_x/B_x},\pm \sqrt{M_z/B_z})$, as long as $\mbox{sign}(M_x B_x) = +1 = \mbox{sign}(M_z B_z)$.
For a fixed $k_z$, $H_{\text{WSM}}$ becomes equivalent to the 2D model studied here, as shown in Fig.~\ref{fig:slices}.

\end{document}